\documentclass[lettersize,journal]{IEEEtran}
\usepackage{amsmath,amsfonts}
\usepackage[linesnumbered,ruled,vlined]{algorithm2e}

\usepackage{array}

\usepackage{textcomp}
\usepackage{stfloats}
\usepackage{url}
\usepackage{verbatim}
\usepackage{amssymb} 
\usepackage{amsfonts}
\usepackage{amsmath} 
\usepackage{physics} 
\usepackage{color} 
\usepackage{bm}
\usepackage{epstopdf} 
\usepackage{subcaption}
\usepackage{graphicx}
\usepackage{booktabs}  
\usepackage[dvipsnames]{xcolor} 
\usepackage{cite} 
\usepackage{balance}
\usepackage{setspace}  
\usepackage{amsthm}
\usepackage{amsmath} 
\usepackage{hyperref}
\usepackage{balance}
\usepackage{placeins} 
\usepackage{scalerel}
\usepackage{caption}

\allowdisplaybreaks 

\newtheorem{ppn}{Proposition}
\newtheorem{lemma}{Lemma} 
\newtheorem{remark}{Remark}

\newtheorem*{pf}{Proof}  

\makeatletter
\renewenvironment{pf}{%
  \par
  \pushQED{\qed}%
  \normalfont \topsep6\p@\@plus6\p@\relax
  \trivlist
  \item[\hskip\labelsep\bfseries Proof.\ ]
  \itshape
  \ignorespaces
}{%
  \popQED
  \endtrivlist
}
\makeatother

\newcommand{\lmref}[1]{\textbf{Lemma \ref{#1}}}

\newcommand{\algref}[1]{\textbf{Algorithm \ref{#1}}}

\newcommand{\figref}[1]{Fig. \ref{#1}}
\newcommand{\secref}[1]{Section \ref{#1}}

\definecolor{softred}{RGB}{200, 50, 50}

\DeclareMathAlphabet{\mathsfit}{\encodingdefault}{\sfdefault}{m}{sl}
\SetMathAlphabet{\mathsfit}{bold}{\encodingdefault}{\sfdefault}{bx}{n}

\begin{document}

\bstctlcite{IEEEexample:BSTcontrol}

\title{Multi-Satellite Multi-Stream\\ Beamspace Massive MIMO Transmission}    
	
\author{Yafei Wang, \textit{Graduate Student Member}, \textit{IEEE},  Yiming Zhu, \textit{Graduate Student Member}, \textit{IEEE}, \\Vu Nguyen Ha, \textit{Senior Member}, \textit{IEEE}, Wenjin Wang, \textit{Member}, \textit{IEEE}, Rui Ding, \\Symeon Chatzinotas, \textit{Fellow}, \textit{IEEE}, Björn Ottersten, \textit{Fellow}, \textit{IEEE} 
\thanks{Manuscript received xxx.}
\thanks{Yafei Wang, Yiming Zhu, and Wenjin Wang are with the National Mobile Communications Research Laboratory, Southeast University, Nanjing 210096, China, and also with Purple Mountain Laboratories, Nanjing 211100, China (E-mail: \{wangyf, ymzhu, wangwj\}@seu.edu.cn).}
\thanks{Vu Nguyen Ha, Symeon Chatzinotas, and Björn Ottersten are with the Interdisciplinary Centre for Security, Reliability and Trust (SnT), University of Luxembourg (E-mails: \{vu-nguyen.ha, symeon.chatzinotas, bjorn.ottersten\}@uni.lu).}
\thanks{Rui Ding is with China Satellite Network Group Company Ltd., Beijing 100029, China (e-mail: greatdn@qq.com).}}
		
    \markboth{}%
    {Shell \MakeLowercase{\textit{et al.}}: A Sample Article Using IEEEtran.cls for IEEE Journals}
    
    \maketitle

\begin{abstract}

This paper studies multi-satellite multi-stream (MSMS) beamspace transmission, where multiple satellites cooperate to form a distributed multiple-input multiple-output (MIMO) system and jointly deliver multiple data streams to multi-antenna user terminals (UTs), and beamspace transmission combines earth-moving beamforming with beam-domain precoding. 
For the first time, we formulate the signal model for MSMS beamspace MIMO transmission. Under synchronization errors, multi-antenna UTs enable the distributed MIMO channel to exhibit higher rank, supporting multiple data streams. Beamspace MIMO retains conventional codebook based beamforming while providing the performance gains of precoding. Based on the signal model, we propose statistical channel state information (sCSI)-based optimization of satellite clustering, beam selection, and transmit precoding, using a sum-rate upper-bound approximation. With given satellite clustering and beam selection, we cast precoder design as an equivalent covariance decomposition-based weighted minimum mean square error (CDWMMSE) problem. To obtain tractable algorithms, we develop a closed-form covariance decomposition required by CDWMMSE and derive an iterative MSMS beam-domain precoder under sCSI. Following this, we further propose several heuristic closed-form precoders to avoid iterative cost. For satellite clustering, we enhance a competition-based algorithm by introducing a mechanism to regulate the number of satellites serving certain UT. Furthermore, we design a two-stage low-complexity beam selection algorithm focused on enhancing the effective channel power. Simulations under practical configurations validate the proposed methods across the number of data streams, receive antennas, serving satellites, and active beams, and show that beamspace transmission approaches conventional MIMO performance at lower complexity.
\end{abstract}
	

	%
	\IEEEpeerreviewmaketitle
	
\vspace{-3mm}	
	
\section{Introduction}

\vspace{-1mm}


    \IEEEPARstart{S}{ixth}-generation (6G) wireless networks identify satellite communication (SatCom) as a critical technology to overcome the coverage limitations of cellular networks, enabling seamless global connectivity as a core vision of IMT-2030 \cite{10820534, WANG2025, cao2025interference}. As a key technique for achieving high spectral efficiency, precoding was studied early in SatCom with DVB-S2X and also gained attention in broadband transmission \cite{3GPP_TR_38_811, 3gpp_tr_38_821}. By modulating the phase and amplitude of radio frequency (RF) signals derived from user data, precoding leverages multi-antenna systems to suppress multi-user interference, enabling the transmission of multiple data streams within the same time-frequency resource block, thus enabling spatial multiplexing and enhancing overall system capacity \cite{bjornson2014optimal, wang2022weighted}. However, challenges in SatCom, such as long propagation delays and high Doppler shifts, hinder instantaneous channel state information (CSI) acquisition, complicating the design of precoding schemes that rely on accurate CSI \cite{11049893}. Practical SatCom systems commonly adopt codebook-based beamforming by selecting beam codewords from a predefined codebook. In contrast, dynamic digital precoding instead introduces much higher computational complexity and a larger number of RF chains, and both grow with the number of transmit antennas, thereby increasing cost and limiting deployment.

    
    Due to the difficulty of acquiring instantaneous CSI, designing precoding based on long-term statistical CSI (sCSI), such as angles of departure and arrival, is a more practical and robust scheme that reduces channel estimation overhead and feedback requirements \cite{you2020massive, li2021downlink, WANG2025, Dong2025statistical}. As multi-satellite cooperation in mega constellations demonstrates significant gains, transmission design in multi-satellite distributed multiple-input multiple-output (MIMO) systems has become a promising technology as it can drastically improve key performance metrics such as signal-to-interference-plus-noise ratio (SINR) and achievable sum rate. Well-crafted downlink strategies are therefore essential to fully exploit the potential of 6G SatCom systems. In addition, beamspace MIMO architectures that integrate codebook-based beamforming and beam-domain precoding have recently attracted attention in SatComs \cite{10008605, ha2024user, Dong2025statistical}, offering an excellent performance and complexity tradeoff while remaining compatible with existing system designs. Motivated by these developments, this work investigates sCSI-based transmission design for distributed beamspace MIMO in orthogonal frequency division multiplexing (OFDM)-based SatComs with multi-antenna user terminals (UTs), and presents theoretical analyses and solutions that support the development of ubiquitous connectivity in 6G.
    
\vspace{-3mm}

\subsection{Related Works}   

Previous studies have extensively examined various precoding techniques for single-satellite systems. The authors of \cite{8353925} investigated minimum mean square error (MMSE)-based precoding, scheduling, and link adaptation, and analyzed the impact of outdated CSI. To enhance robustness against phase errors and CSI distortion, \cite{8629918} and \cite{wang2021resource} proposed precoding algorithms for power minimization and resource efficiency maximization. Improving transmission performance by enlarging the antenna array is another important direction. In particular, \cite{you2020massive} analyzed the channel characteristics and designed an sCSI-based precoding scheme that improves average signal-to-leakage-plus-noise ratio (SLNR) performance. In addition, \cite{10437228} introduced a low-complexity precoding update algorithm that mitigates performance degradation caused by CSI aging. However, increasing the number of satellite antennas significantly raises computational complexity and implementation cost, making conventional MIMO precoding far less practical than codebook-based beamforming in existing SatCom systems. To this end, \cite{10008605} proposed an architecture that integrates the two approaches and resembles beamspace transmission \cite{gao2016near, wu2023simultaneous}. It preserves the codebook-based beamforming hardware while introducing precoding in the beam domain, achieving an excellent balance between performance and complexity. Building on this architecture, \cite{Dong2025statistical} further enabled flexible adjustment of the number of active beams and developed new sCSI-based transmission schemes.

Given the inherent limitations of current satellite payloads and antenna manufacturing, the communication capacity of an individual satellite remains fundamentally constrained. Therefore, multi-satellite distributed MIMO transmission achieved through inter-satellite cooperation has become a promising solution for dense low Earth orbit (LEO) satellite constellations, as they efficiently aggregate communication resources from multiple satellites. Specifically, \cite{9939157} proposed a cell-free LEO satellite framework and developed two joint power allocation and handover schemes that mitigate inter-satellite interference and improve service continuity.
\cite{10440321} studied multi-satellite noncoherent joint transmission (NCJT), where UTs receive distinct data streams from different satellites and the streams interfere asynchronously \cite{liu2025gcn}. However, since multiple satellites cannot deliver common streams, it is mainly suitable for UTs with many antennas. In contrast, multi-satellite coherent joint transmission (CJT) enables satellites to transmit the same synchronized stream, substantially improving link quality and system capacity and supporting UTs with few antennas, even a single antenna, which has attracted broad attention.
For example, \cite{10380500} introduces a hybrid precoding architecture for multi-satellite CJT.
Unlike most centralized designs, \cite{zhang2025decentralized} studied a decentralized iterative optimization for sCSI-based CJT.
Synchronization, including delay and phase alignment, is a key factor affecting CJT performance.
Focusing on the impact of asynchrony under the DVB-S2X standard, \cite{10596023} designed asynchronous weighted MMSE (WMMSE) and delay estimation algorithms using instantaneous CSI and timing-error information. For OFDM systems, \cite{wu2025distributed} analyzed the effect of synchronization errors on distributed beamforming. Recent works \cite{wang2025DP} and \cite{wang2025DP_Conf} further investigated how 
synchronization errors introduce new characteristics into the OFDM signal model and developed novel optimization and deep learning-based precoders for single-antenna UTs. 
Furthermore, in the context of beamspace MIMO transmission, \cite{ha2024user} extended the framework of \cite{10008605} to a multi-satellite network and devised tailored algorithms, but each UT is served by only one satellite simultaneously.

\vspace{-4mm}
\subsection{Contributions}
\vspace{-1mm}
    To achieve the ubiquitous connectivity envisioned for 6G, it is crucial to develop multi-satellite distributed MIMO transmission schemes that comply with standards and compatible with practical implementations. However, most existing studies on multi-satellite coherent distributed MIMO focus on single-antenna or single data stream, and the multi-satellite channel characteristics of CJT associated with multi-antenna UTs specified in current standards have not been fully revealed or exploited \cite{3GPP_TR_38_811, 3gpp_tr_38_821}. Moreover, although beamspace transmission has demonstrated outstanding performance and complexity tradeoff advantages in SatComs, its design under sCSI for multi-satellite distributed MIMO remains insufficiently explored. Given its practical relevance, sCSI-based transmission with synchronization errors deserves further attention. These observations highlight a research problem of both theoretical and practical significance: \textit{How to design multi-satellite multi-stream transmission for distributed beamspace MIMO systems to further advance ubiquitous connectivity in 6G?} This work addresses this challenge by proposing a new multi-satellite multi-stream (MSMS) CJT framework that integrates satellite clustering, beam selection, and beam-domain transmit precoding. The major contributions of this work are as follows:
 \begin{itemize}
    \item Under OFDM modulation, we firstly propose a multi-satellite distributed MU-MIMO channel model that accounts for inter-satellite synchronization errors with multi-antenna UTs. We then extend such system by considering beamspace MIMO transmission, which combines beam selection with distributed precoding over the equivalent beamspace channel. This design is well matched to practical SatComs with beamforming and line-of-sight (LoS)-dominant channels, and it reduces to conventional MIMO transmission as a special case. On this basis, we formulate an optimization problem that designs satellite clustering, beam selection, and distributed precoding to maximize the weighted ergodic sum rate. The proposed architecture aggregates the spatial resources of multiple satellites while reducing processing complexity, thereby achieving an attractive performance–complexity tradeoff.
    \item For the joint optimization problem, we adopt a two-stage approach: we first optimize beam selection and then design distributed precoding over the resulting beamspace equivalent channels. For MSMS distributed precoding, we exploit sCSI that incorporates synchronization-error statistics and approximate the sum rate objective by an upper bound, thereby avoiding the pronounced underestimation of effective signal power inherent in the lower-bound approximations commonly used in existing work. We then reformulate the problem into an equivalent covariance decomposition-based optimization and develop a dedicated covariance decomposition method. Building on this formulation, we propose an iterative MSMS distributed precoding algorithm. To further enhance the performance-complexity tradeoff, we also propose two closed-form sCSI-based precoding schemes. 
    \item 
    For satellite clustering, we adopt a competition-based algorithm driven by sCSI, while designing a redundancy-removal module to prevent the number of serving satellites from exceeding the limit. As the foundation of beamspace MIMO, we exploit inherent characteristics of SatCom to design a low-complexity sCSI-based beam selection algorithm whose computational complexity is independent of the number of antennas and centers on enhancing the effective channel power. 
    \item 
    Based on practical deployment settings, we conducted comprehensive evaluations over scenarios and channels generated by an authoritative channel simulator. The experiments assessed the proposed scheme across multiple dimensions, including the number of streams, receive antennas, serving satellites, and active beams, and confirmed its performance gains. Moreover, beamspace MIMO transmission exhibits the ability to approach conventional MIMO performance with significantly lower complexity, highlighting its practical value.
\end{itemize}


This paper is structured as follows: The system model, signal model, and optimization problem are presented in \secref{signal model sec}. Iterative precoding algorithms are investigated in \secref{CDWM sec}. Two closed-form precoders are proposed in \secref{CF precoding sec}. Satellite clustering and beam selection algorithms are examined in \secref{scheduling sec}. Simulation results are reported in \secref{result sec}, and conclusions are drawn in \secref{conclusion sec}.

	{\textit{Notation}}: $x, {\bf x}$, and ${\bf X}$ represent scalar, column vector, and matrix. $(\cdot)^T$, $(\cdot)^{*}$, $(\cdot)^H$, and $(\cdot)^{-1}$ denote the transpose, conjugate, transpose-conjugate, and inverse operations, respectively. ${\bf I}_{M}$ represents $M\times M$ identity matrix. $\left \|\cdot\right \|_{2}$ denotes $\ell_2$-norm. $\otimes$ is the Kronecker product operations. The operator ${\rm Tr}(\cdot)$ represents the matrix trace.  ${{\rm diag}\{{\bf a}\}}$ represents a diagonal matrix whose diagonal elements are composed of ${\bf a}$. The expression $\mathcal{C}\mathcal{N}(\mu, \sigma^2)$ denotes circularly symmetric Gaussian distribution with expectation $\mu$ and variance $\sigma^2$. ${\mathbb{R}}^{M\times N}$ and ${\mathbb{C}}^{M\times N}$ represent the set of $M\times N$ dimension real- and complex-valued matrixes. $\nabla f$ denotes gradient of function $f(\cdot)$. $k\in \mathcal{K}$ means element $k$ belongs to set $\mathcal{K}$.

    \vspace{-2mm}
    \section{System Model \& Problem Formulation}
    \label{signal model sec}
    \vspace{-1mm}

    \begin{figure}[!t]
        \centering
        \includegraphics[width=3in]{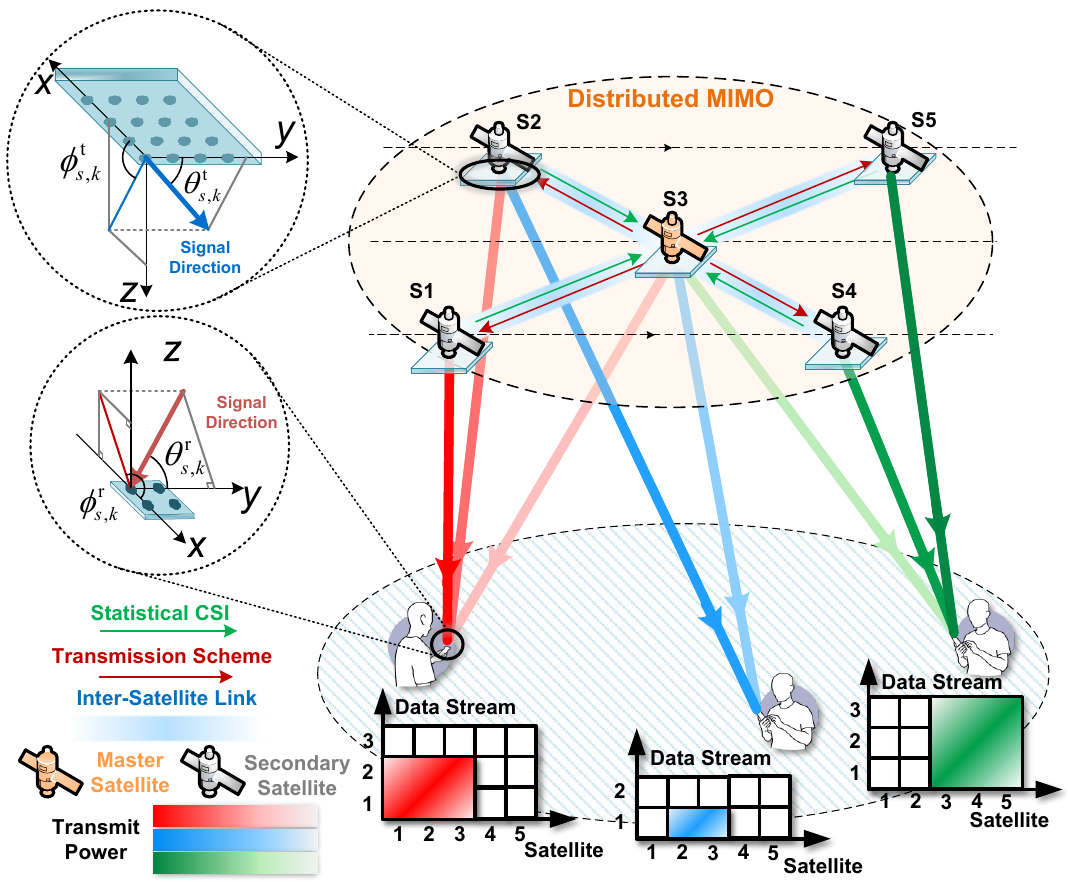}
        \captionsetup{font=footnotesize}
        \caption{Illustration of the MSMS distributed massive MIMO system.}
        \label{system model fig}
        \vspace{-6mm}
    \end{figure}

    As shown in \figref{system model fig}, we consider the downlink of MSCT system, where $K$ UTs are served by $S$ satellites over the same time-frequency resources. 
    Satellites and UTs employ uniform planar arrays (UPAs) with $N_{\rm T}=N_{\rm TV}N_{\rm TH}$ and $N_{\rm R}=N_{\rm RV}N_{\rm RH}$ antennas, respectively.
    The gateway stations (GSs) sends user data to the satellites individually, which increases feeder-link demand and can be met with advanced GS technologies such as multi-antenna arrays \cite{liu2025time, 9625524}.
    On the time-frequency resource of interest, the set of all UTs is denoted by $\mathcal{K}$ ($|\mathcal{K}|=K$), and the set of all cooperative satellites is denoted by $\mathcal{S}$  ($|\mathcal{S}|=S$). From a user-centric perspective, the set of satellites serving UT $k$ is denoted by $\mathcal{S}_k\subseteq \mathcal{S}$. As illustrated in the tables of \figref{system model fig}, each UT is served by multiple satellites in the set $\mathcal{S}_k$ to transmit one or multiple data streams.
    The secondary satellites forward low-dimensional sCSI to the master satellite via inter-satellite links (ISLs), which computes transmission schemes and distributes them with related signaling over the ISLs.

    \vspace{-3mm}	
    \subsection{Multi-Satellite Channel Model}
    \label{channel sec}
    \vspace{-1mm}	
    
    According to \cite{li2021downlink, hou2024joint, zhu2024joint}, the time-varying spatial domain MIMO channel ${\tilde {\bf H}}_{s,k}(t, \tau)\in\mathbb{C}^{N_{\rm R}\times N_{\rm T}}$ between the $s$-th satellite and the $k$-th UT is given by
    \vspace{-2mm}
    \begin{align}
    {\tilde {\bf H}}_{s,k}(t, \tau) \!=\! \sum_{l=1}^{L_{s,k}}a_{s,k,l}\cdot{\rm e}^{j2\pi \nu_{s,k,l}t}\delta(\tau\!-\!\tau_{s,k,l}){\bf u}_{s,k,l}{\bf v}^T_{s,k},
    \label{delay channel eq}
    \end{align}
    where $t$ and $\tau$ represent time and delay; $l$ is the index of the path, and the total number of paths in the channel is $ L_{s, k} $; $a_{s,k,l}$, $\nu_{s,k,l}$, and $\tau_{s,k,l}$ represent the complex gain, Doppler frequency offset, and delay of the path channel, respectively; ${\bf u}_{s,k,l}\in\mathbb{C}^{N_{\rm R}\times 1}$ and ${\bf v}_{s,k}\in\mathbb{C}^{N_{\rm T}\times 1}$ are the steering vectors at the receiver and transmitter.

    For multi-satellite coherent distributed MIMO, multiple satellites perform satellite-side pre-compensation for delay and Doppler shifts with respect to the UT, thereby enabling effective superposition of signals from multiple satellites at the receiver \cite{wang2025DP, wu2025distributed,liu2025gcn}. The foundation of distributed MIMO lies in the assumption that the arrival times of signals from multiple satellites are aligned within the cyclic prefix of one OFDM symbol. This alignment can be achieved by estimating the propagation delay with global navigation satellite system (GNSS) or downlink pilots and predicting future delays based on trajectory information. In this work, we assume imperfect delay and Doppler compensation, while the delay still satisfies the above basic assumption and the Doppler error is much smaller than the subcarrier spacing.    
    As in most prior studies \cite{10596023,wu2025distributed,wang2025DP}, we assume synchronized local oscillators across satellites. After pre-compensation, the received signal at UT $k$ on the $n$-th subcarrier and $m$-th OFDM symbol is \cite{wu2025distributed, wang2025DP}
    \vspace{-1mm}
    \begin{align}
     \begin{split}
     &\textstyle{\bf y}^{(m)}_{k,n} \!=\! \sum_{s\in\mathcal{S}_{k}}\!{\bf H}^{(m)}_{s,k,n} {\bf x}^{(m)}_{s,k,n}\varphi^n_{s,k,k} \\
     &\textstyle\qquad \qquad + \sum_{j\neq k}\sum_{i\in\mathcal{S}_j}{\bf H}^{(m)}_{i,k,n} {\bar {\bf x}}^{(m'_{i,j,k})}_{i,j,n}\varphi^n_{i,j,k} + {\bf n}^{(m)}_{k,n},
     \end{split}
     \label{received signal eq1}
    \end{align}
    where ${\bf x}^{(m)}_{s, k, n}\in\mathbb{C}^{N_{\rm T}\times 1}$ is the precoded desired signal transmitted from satellite $s$ to UT $k$. 
    where ${\bar {\bf x}}^{(m'_{i,j,k})}_{i,j,n}$ denotes the portion of the signal transmitted from satellite $i$ to UT $j$ that impacts UT $k$, exhibiting inter-carrier interference (ICI) and inter-symbol interference (ISI) due to pre-compensation effects \cite{zhu2025downlink}. 
    The superscript $m'_{i,j,k}$ indicates the range of symbol indices spanned under ISI, 
which is determined jointly by the delay pre-compensation from satellite $i$ to UT $j$ 
and the channel delay from satellite $i$ to UT $k$. ${\bf n}^{(m)}_k\in\mathbb{C}^{N_{\rm R}\times 1}$ denotes the additive white Gaussian noise vector
with distribution ${\mathcal{CN}}({\bf 0}, \sigma^2_k{\bf I})$. $\varphi^n_{s,k,k}$ is the phase error introduced by the pre-compensation error, and $\varphi^n_{i,j,k}$ is the phase carried by the interference signal and is influenced by the pre-compensation and the channel. The unified expression for both is \cite{wang2025DP}
    \begin{align}
        \varphi^n_{s,j,k} = {\rm e}^{j2\pi (f_0+n\Delta f-\nu_{s,j}^{\rm cps})(\tau_{s,j}^{\rm cps}-\tau_{s,k})},
    \end{align}
    where $\nu_{s,j}^{\rm cps}$ and $\tau_{s,j}^{\rm cps}$ are the Doppler and delay pre-compensations from satellite $s$ to UT $j$, respectively; $ f_0 $ is the carrier frequency. $ \Delta f $ is the subcarrier spacing, and $\tau_{s,k}\triangleq\tau_{s,k,1}$ represents the delay of the LoS path. 
    Due to the large magnitude of the carrier frequency $f_0$, eliminating this phase error via pre-compensation is highly challenging.
    In \eqref{received signal eq1}, ${\bf H}^{(m)}_{s, k,n}\in\mathbb{C}^{N_{\rm R}\times N_{\rm T}}$ denotes the channel frequency response dominated by the LoS path after synchronization. Without loss of generality, its expression after omitting the subcarrier and symbol indices is given below \cite{li2021downlink}, \cite{wu2025distributed}
    \begin{align}
        {\bf H}_{s,k} &\textstyle= \sqrt{\frac{\kappa_{s,k}\gamma_{s,k}}{\kappa_{s,k} + 1}}{\bf H}^{\rm LoS}_{s,k} + \sqrt{\frac{\kappa_{s,k}}{\kappa_{s,k} + 1}}{\bf H}^{\rm NLoS}_{s,k}\\
        &\textstyle= \left(\sqrt{\frac{\kappa_{s,k}\gamma_{s,k}}{\kappa_{s,k} + 1}}{\bf u}_{s,k} + \sqrt{\frac{\kappa_{s,k}}{\kappa_{s,k} + 1}}{\tilde {\bf u}}_{s,k}\right){\bf v}^T_{s,k} \\
        &\textstyle= {\bar {\bf u}}_{s,k}{\bf v}^T_{s,k}, \label{frequency channel model eq}
    \end{align}
    where $\gamma_{s,k} = {\mathbb E}\{{\rm Tr}({\bf H}_{s,k}{\bf H}^H_{s,k})\}$ represents the average channel power, and $\kappa_{s,k}$ denotes the Rician factor. ${\bf H}^{\rm NLoS}_{s,k}={\tilde {\bf u}}_{s,k}{\bf v}^T_{s,k}$ is the random non-line-of-sight (NLoS) channel introduced by scatterers around the UT, characterizing the NLoS component in \eqref{delay channel eq}, where ${\tilde {\bf u}}_{s,k}\sim{\mathcal {CN}}({\bf 0}, {\boldsymbol{\Sigma}}_{s,k})$. ${\bf H}^{\rm LoS}_{s,k}={\bf u}_{s,k}{\bf v}^T_{s,k}$ denotes the dominant line-of-sight (LoS) path channel, where the expressions for the steering vectors ${\bf v}_{s,k}$ and ${\bf u}_{s,k}\triangleq {\bf u}_{s,k,1}$ are expressed as
    \vspace{-2mm}
    \begin{align}
        {\bf v}_{s,k} &= {\bf v}_{N_{\rm TV}}(\cos ({\theta^{\rm t}_{s,k}}))\otimes  {\bf v}_{N_{\rm TH}}(\sin(\smash{\theta^{\rm t}_{s,k}})\cos(\phi^{\rm t}_{s,k})),\label{eq v}\\
        {\bf u}_{s,k} &= {\bf v}_{N_{\rm RV}}(\cos ({\theta^{\rm r}_{s,k}}))\otimes  {\bf v}_{N_{\rm RH}}(\sin(\smash{\theta^{\rm r}_{s,k}})\cos(\phi^{\rm r}_{s,k})),\label{eq u}
    \end{align}
    where ${\phi^{\rm t}_{s,k}}$ and ${\theta^{\rm t}_{s,k}}$ are the departure azimuth and elevation angles of the signal, respectively; ${\phi^{\rm r}_{s,k}}$ and ${\theta^{\rm r}_{s,k}}$ are the arrival azimuth and elevation angles of the signal, with their specific definitions illustrated in \figref{system model fig}. For simplicity of notation, we define ${\boldsymbol{\theta}}_{s,k}\triangleq[{\theta^{\rm t}_{s,k}},{\phi^{\rm t}_{s,k}}, {\theta^{\rm r}_{s,k}},{\phi^{\rm r}_{s,k}}]^T$. The vector ${\bf v}_{N}(x)\in\mathbb{C}^{N\times 1}$ is defined as $\textstyle {\bf v}_{N}(x) = \frac{1}{\sqrt{N}}\cdot [{\rm e}^{-j\pi0x}, {\rm e}^{-j\pi1x}, ..., {\rm e}^{-j\pi(N-1)x}]$.

    In models \eqref{received signal eq1} and \eqref{frequency channel model eq}, the following aspects are noteworthy:
    \vspace{-4mm}
    \begin{itemize}
        \item{\textit{Synchronization Error:}} After achieving symbol-level synchronization, the residual time-frequency pre-compensation error is transformed into a hard-to-eliminate phase $\varphi^n_{s,k,k}$ acting on the LoS path \cite{wang2025DP, wu2025distributed}. In addition, this phase can also encompass all other phase errors, such as those introduced by imperfect hardware conditions. We assume that this phase follows a random distribution with mean ${\bar \varphi}_{s,k} = \mathbb{E}\{\varphi^n_{s,k,k}\}$.
        Since this phase depends on multiple factors, including compensation accuracy, attitude and velocity, and hardware imperfections, we assume that these phase errors are independently distributed.
        \item{\textit{Asynchronous Interference:}} Pre-compensation induces time and frequency offsets in the interference. According to \cite{wang2025DP}, for typical co-orbital or non-co-orbital satellites, the time offset differences among interferences from different satellites to the same UT exceed one OFDM symbol duration, rendering these interference signals (e.g., ${\bar {\bf x}}^{(m'_{i_1,j,k})}_{i_1,j,n}$ and ${\bar {\bf x}}^{(m'_{i_2,j,k})}_{i_2,j,n}$) independent and thus uncorrelated. 
        Although satellites in a swarm may be close enough to maintain OFDM symbol alignment \cite{de2025applicability,Tamiru2025Distributed}, the presence of the phase $\varphi^n_{i,j,k}$ may still reduce the correlation among interference signals. Furthermore, although frequency offsets generate ICI dispersing across all subcarriers \cite{zhu2025downlink}, filtered-OFDM (F-OFDM) can be employed to mitigate such dispersion of interference.
        \item{\textit{Multi-Rank Structure:}} According to \eqref{frequency channel model eq}, despite multiple antennas at the receiver, the channel from a single satellite to a single UT remains rank-one. It is straightforward to prove that transmitting a single spatial stream already achieves optimal performance. In contrast, the multi-satellite channel ${\bf H}^{\rm MS}_{k} = [{\bf H}_{s, k}]_{2, s\in\mathcal{S}_k}\in\mathbb{C}^{N_{\rm R}\times S_kN_{\rm T}}$, where $[\cdot]_2$ denotes column-wise concatenation, exhibits a multi-rank structure, which further enables the design of multi-stream transmission. 
    \end{itemize}
    Due to the high mobility of satellites, Real-time estimation of the accurate CSI ${\bf H}_{s,k}$ is not practical, and the dominance of the LoS component in channels makes the sCSI framework a feasible approach. 
    As shown in \figref{system model fig}, in the considered system, sCSI ${\mathcal H}_{\rm MS}=\{\gamma_{s,k},\kappa_{s,k},{\boldsymbol{\theta}}_{s,k},{\boldsymbol{\Sigma}}_{s,k},{\bar \varphi}_{s,k}\}_{\forall s,k}$ can be estimated at each satellite and aggregated at the master satellite via ISLs \cite{wang2025DP,li2021downlink}.

    \vspace{-2mm}	
    \subsection{Multi-Satellite Beamspace MIMO Transmission}
    \vspace{-1mm}	
    
    Consistent with most practical SatCom systems, we assume that all satellites employ earth-moving beamforming, i.e., using beams from a pre-designed beam codebook to direct energy toward a specific relative direction of the satellite.
    Introducing beamspace transmission, which combines beamforming and beam-domain precoding, is a suitable choice. 
    On the one hand, precoding compensates for the inherent limitations of earth-moving beams in interference suppression and multi-satellite phase alignment for power enhancement. On the other hand, the LoS path-dominant nature of satellite channels yields a highly sparse beamspace channel, reducing its effective dimension and processing complexity.

    The selectable beam set for satellite $s$ is denoted as  $\mathcal{Q}_s$, with the number of beams (i.e., codebook size) being $|\mathcal{Q}_s|=Q_s$, while the set of activated beams by satellite $s$ is $ \mathcal{B}_s$ ($|\mathcal{B}_s| = B_s$). Ignoring the subcarrier and symbol indices, the received signal model \eqref{received signal eq1} can be expressed as
    \vspace{-2mm}
    \begin{align}
        \begin{split}
            {\bf y}_{k} &\textstyle= \sum_{s\in\mathcal{S}_{k}}\!{\bf H}_{s,k}{\bf F}_s{\bf A}_s {\bf W}_{s,k}{\bf d}_{k}\varphi_{s,k,k}
        \\
        &\textstyle\quad+ \sum_{j\neq k}\sum_{i\in\mathcal{S}_j}{\bf H}_{i,k} {\bf F}_i{\bf A}_i{\bf W}_{i,j}{\bar {\bf d}}^{i,k}_{j}\varphi_{i,j,k} + {\bf n}_{k},
        \end{split}
        \label{received signal model eq2}
    \end{align}
    where ${\bf F}_{s}\in\mathbb{C}^{N_{\rm T}\times Q_s}$ denotes the beam codebook for satellite $s$, and ${\bf A}_{s}\in\{0, 1\}^{Q_s\times B_s}$ is the beam selection matrix.
    Under the distributed MIMO architecture, cooperation is carried out in a CJT manner \cite{liu2025gcn, 3GPP_TR_36_819}, where the antennas of satellites jointly form a larger distributed array to transmit the same data to given UTs. In this framework, ${\bf d}_k\in\mathbb{C}^{M_k\times 1} $ represents the transmitted information for UT $k$, ${\bf W}_{s,k}\in\mathbb{C}^{B_s\times M_k}$ is the precoding from satellite $s$ to UT $k$, and $M_k$ is the number of data streams. The selection range for $M_k$ is as follows:
    \vspace{-2mm}
    \begin{remark}
        CJT enables multiple satellites to convey identical data streams, yielding a minimum $M_k$ of 1. Besides limitations from receiver antenna capability, the maximum $M_k$ is constrained by the number of serving satellites due to the rank-one nature of single-satellite channels. Thus, the number of data streams satisfies $1\leq M_k \leq {\rm min}\{|\mathcal{S}_k|,N_{\rm R}\}$.
    \end{remark}
    \vspace{-2mm}
    From the precoding perspective, the beam codebook and beam selection, along with the original channel, form the multi-satellite beamspace channel. Thus, equation \eqref{received signal model eq2} can be rewritten as follows:
    \begin{align}
        \begin{split}
            &\textstyle{\bf y}_{k} =\!\sum_{s\in\mathcal{S}_{k}}\!{\bar {\bf H}}_{s,k}{\bf W}_{s,k}{\bf d}_{k}\varphi_{s,k,k}+\\
            &\textstyle\qquad\qquad\sum_{j\neq k}\!\sum_{i\in\mathcal{S}_j}\!{\bar {\bf H}}_{i,k}{\bf W}_{i,j}{\bar {\bf d}}_{j}\varphi_{i,j,k} \!+ \!{\bf n}_{k},
        \end{split}
    \end{align}
    where ${\bar {\bf H}}_{s,k}\triangleq{\bf H}_{s,k}{\bf F}_s{\bf A}_s\in\mathbb{C}^{N_{\rm R}\times  B_s}$. 
    Based on this, stacking the serving satellites' channels column-wise yields the multi-satellite beam-domain channel ${\bar {\bf H}}^{\rm MS}_{k}=[{\bar {\bf H}}_{s,k}]_{2,s\in\mathcal{S}_k}\in\mathbb{C}^{N_{\rm R}\times \sum_{s\in\mathcal{S}_k} B_s}$.
    From an equivalent channel perspective, the matrices ${\bf F}_s$ and ${\bf A}_s$ perform beam-domain sampling on the original channel and cropping of the beamspace channel, respectively. Discrete Fourier transform (DFT) beams are employed as the beamforming codewords, i.e., the matrix $ \mathbf{F}_s \in\mathbb{C}^{N_{\rm T}\times Q_s}$ is given by ${\bf F}_{s} = ({\tilde {\bf F}}_{N_{\rm TV}}\otimes {\tilde {\bf F}}_{N_{\rm TH}} )^{*}$, where $\textstyle{\tilde {\bf F}}_{N[:, n]} = {\bf v}_{N}(-1+\frac{2(n-1)}{N})$ and $Q_s=N_{\rm T}$.
    From the receiver perspective, multi-satellite distributed MIMO is nearly equivalent to conventional MIMO, and multi-stream reception can therefore follow MIMO techniques \cite{tse2005fundamentals, liu2025gcn}.

    \vspace{-2mm}
    \subsection{Problem Formulation}
    \vspace{-2mm}
    For beamspace transmission, beam selection and beam-domain transmit precoding are essential optimization variables. In addition, user-centric satellite clustering, namely the selection of serving satellites, is also a key design aspect and can be used to reduce the feeder link burden. 
    For convenience in subsequent expressions, we first define the matrix ${\bf O}\in\{0, 1\}^{S\times K}$, whose element $o_{s,k}$ is the indicator for the user-satellite association. It takes values of 1 or 0, indicating whether satellite $s$ serves UT $k$ or not, respectively. Therefore, ${\mathcal S}_k = \{s|o_{s,k}=1,s\in\mathcal{S}\}$. 
    In the system, UT $k$ is served by $|\mathcal{S}_k|=S_k$ satellites. Satellite $s$ serves $K_s$ UTs simultaneously, with $K_s \le K_s^{\rm max}$. In summary, the optimization problem is formulated as follows:
    \begin{align}
    \begin{split}
    ({\text P1}): &\max\limits_{\{{\bf W}_{s,k}\}, \{{\bf A}_s\}, {\bf O}}\ {\sum_{\forall k}}\beta_kR_k\\
    {\rm s.t.}\ 
    &\textstyle ({\rm C1})\  \sum_{s\in\mathcal{S}} o_{s, k}=S_{k},\ \forall k\in\mathcal{K},\\
    &\textstyle ({\rm C2})\  \sum_{k\in\mathcal{K}} o_{s, k}\leq K^{\rm max}_s,\ \forall s\in\mathcal{S},\\
    &\textstyle ({\rm C3})\  \sum_{q=1}^{Q_s} a_{s, q, b}=1,\ \forall b=1,...,B_s,\ s\in\mathcal{S}, \\
    &\textstyle ({\rm C4})\  \sum_{b=1}^{B_s} a_{s, q, b}\leq 1,\ \forall q=1,...,Q_s,\ s\in\mathcal{S}, \\
    &\textstyle ({\rm C5})\  o_{s, k}\in\{0, 1\},\ \forall s, k,\\
    &\textstyle ({\rm C6})\  a_{s, q, b}\in\{0, 1\},\ \forall s,q, b,\\
    &\textstyle ({\rm C7})\  \sum_{k\in\mathcal{K}}{\rm Tr}\left({\bf W}_{s,k}{\bf W}^H_{s,k}\right) \leq P_s,\ s\in\mathcal{S}.
    \label{modified wmmse problem 1}
    \end{split}
    \end{align}
    Constraints (C1), (C2), and (C5) pertain to user-centric satellite clustering, while (C3), (C4), and (C6) relate to beam selection. Constraint (C7) represents the power constraint on beam-domain precoding.
    In the objective function, $\beta_k$ denotes the rate weight for the UT, and the achievable rate $R_k$ is expressed as
    \vspace{-2mm}
    \begin{align}
    R_k =\mathbb{E}_{{\bf H}, \boldsymbol{\varphi}}\left\{\log_2\det\left({\bf I} + {\bf R}^{-1}_{{\rm other}, k}{\bf R}_{{\rm sig}, k}\right)\right\},
\end{align}
where ${\bf R}_{{\rm sig}, k} \!\!=\!\! \mathbb{E}_{{\bf d},{\bf n}}\{{\bf y}_{{\rm sig}, k}{\bf y}^H_{{\rm sig}, k}\}$, ${\bf R}_{{\rm other}, k} \!\!=\!\! \mathbb{E}_{{\bf d},{\bf n}}\{{\bf y}_{{\rm other }, k}{\bf y}^H_{{\rm other }, k}\}$, and
\begin{align}
&\textstyle{\bf y}_{{\rm sig }, k} = \sum\nolimits_{s\in\mathcal{S}_{k}}{\bf H}_{s,k}{\bf F}_s{\bf A}_s {\bf W}_{s,k}{\bf d}_{k}\varphi_{s,k,k},\\
&\textstyle{\bf y}_{{\rm other }, k} = \sum_{j\neq k}\sum\nolimits_{i\in\mathcal{S}_j}{\bf H}_{i,k} {\bf F}_i{\bf A}_i{\bf W}_{i,j}{\bar {\bf d}}_{j}\varphi_{i,j,k} + {\bf n}_{k}.
\end{align}
The information among UTs is mutually independent, and it is assumed that ${\bf d}_k\sim\mathcal{CN}({\bf 0}, {\bf I}_{M_k})$. According to the analysis in \secref{channel sec}, asynchronous interference induces ICI, with the interference dispersed across the entire frequency band. If the UT sets across different frequency domain resources are identical, then ${\bar {\bf d}}_{k}\sim\mathcal{CN}({\bf 0}, {\bf I}_{M_k})$. Otherwise, more refined interference modeling is required, as detailed in \cite{zhu2025downlink}, or F-OFDM can be employed.
Problem (P1) is a mixed-integer optimization problem with a non-convex objective function, making it challenging to solve via conventional optimization methods. In the following, we proceed with a step-by-step analysis and design solution algorithms leveraging unique multi-satellite channel characteristics.

\vspace{-1mm}
\section{CDWMMSE Distributed Precoding Design}
\label{CDWM sec}
\vspace{-1mm}

To effectively address (P1) in a feasible manner, we first focus on precoding design to achieve optimal performance under given beam selection and satellite clustering schemes.

\vspace{-3mm}
\subsection{Upper Bound Approximation}
\vspace{-1mm}

Since $R_k$ admits no simple closed-form expression, we approximate it as
\vspace{-2mm}
\begin{align}
\textstyle
({\text {P2}}):\quad \max\limits_{\{{\bf W}_{s,k}\}}\ {\sum_{\forall k}}\beta_k{\bar R}_k \;
\quad{\rm s.t. \ \  (C7),}
\label{modified wmmse problem 2}
\end{align}
\vspace{-2mm}
where ${\bar R}_k =\log_2\det({\bf I} + {\bar {\bf R}}^{-1}_{{\rm other}, k}{\bar {\bf R}}_{{\rm sig}, k})$ and
\begin{align}
    &{\bar {\bf R}}_{{\rm other}, k} \!=\! \mathbb{E}_{{\bf H}, \boldsymbol{\varphi}}\left\{{{\bf R}}_{{\rm other}, k}\right\},\ {\bar {\bf R}}_{{\rm sig}, k} \!=\! \mathbb{E}_{{\bf H}, \boldsymbol{\varphi}}\left\{{{\bf R}}_{{\rm sig}, k}\right\}.
\end{align}
This expression serves as an upper bound when ${\bar {\bf R}}_{{\rm other}, k}$ is known at UT $k$.
According to \cite{wang2025DP} and \cite{Dong2025statistical}, the upper-bound approximation prevents the issue of significant effective signal power underestimation in the lower-bound approximation due to phase errors, offering improved performance, although the either upper and lower bound approximations for MSMS sum rate have not yet been studied. However, the problem is not equivalent to minimizing MSE, rendering the conventional WMMSE formulation inapplicable \cite{10596023, shi2011iteratively}. The unique sCSI framework further impedes traditional algorithms. Although \cite{wang2025DP} provides an optimization algorithm for single-antenna UTs with single data stream, solving ({\text {P2}}) for multi-antenna UTs and multi-stream transmission remains highly challenging due to differences in channel models and transmission modes.

\vspace{-3mm}
\subsection{CDWMMSE Optimization Problem}
\vspace{-1mm}

To design readily implementable algorithms for solving $({\text {P2}})$, we provide the following equivalent covariance decomposition-based WMMSE (CDWMMSE) problem:



\begin{ppn}\label{ppn CDWMMSE problem}
Problem (P2) is equivalent to the following one in the sense that they share the same optimal $\{{\bf W}_{s,k}\}$.
\vspace{-2mm}
\begin{align}
({\text{P3}}):\ &\min\limits_{\{{\bf W}_{s,k}\},\{{\bf C}_{k}\},\{{\bf D}_{k}\}}\ {\sum_{\forall k}}\beta_k\left[{\rm Tr}({\bf C}_k{\bf E}_k)-\log\det({\bf C}_k)\right]\notag\\
&\textstyle\quad{\rm s.t.}\  \sum_{k\in\mathcal{K}}{\rm Tr}\left({\bf W}_{s,k}{\bf W}^H_{s,k}\right) \leq P_s,\ s\in\mathcal{S},
\label{modified wmmse problem 3}
\end{align}
where ${\bf E}_k\in\mathbb{C}^{L_k\times L_k}$ is given by
\begin{align}
   \!\!\! {\bf E}_k\! =\!{\bf D}^H_k\!({\bar {\bf R}}_{{\rm sig}, k}\!\!+\!{\bar {\bf R}}_{{\rm other}, k}\!){\bf D}_k\!\!-\!{\bf D}^H_k\!{\bar {\bf R}}_{{\rm sig}, k}^{\frac{1}{2}}\!\!-\!{\bar {\bf R}}_{{\rm sig}, k}^{\frac{1}{2} H}{\bf D}_k\!\! +\!{\bf I}.\!\!
\label{MSE loss function}
\end{align}
${\bar {\bf R}}_{{\rm sig}, k}^{\frac{1}{2}}\in\mathbb{C}^{N_{\rm R}\times L_k}$ is obtained through the covariance matrix decomposition that needs to be designed, i.e., 
\vspace{-2mm}
\begin{align}
    {\bar {\bf R}}_{{\rm sig}, k}={\bar {\bf R}}_{{\rm sig}, k}^{\frac{1}{2}}({\bar {\bf R}}_{{\rm sig}, k}^{\frac{1}{2}})^H,\ \forall k\in\mathcal{K}.
\end{align}
The size of $ L_k $ is determined by the covariance matrix decomposition method and influences the dimensions of ${\bf E}_k$ and the auxiliary variables ${\bf C}_k\in\mathbb{C}^{L_k\times L_k}$ and ${\bf D}_k\in\mathbb{C}^{N_{\rm R}\times L_k}$.
\end{ppn}
\begin{pf}
See Appendix \ref{pf CDWMMSE problem}.\qedhere
\end{pf}

\begin{figure*}[!b]
    \small
    \vspace*{-4.5mm}
    \hrulefill
    \vspace{-2mm}
    {\setlength{\jot}{0.5pt}
    \begin{align}
        {\bar {\bf R}}_{{\rm sig}, k} &\textstyle= \mathbb{E}_{{\bf H}, {\bf d}, \boldsymbol{\varphi}}\left\{\left(\sum_{s_1\in\mathcal{S}}{\bar {\bf H}}_{s_1, k}{\bf W}_{s_1, k}{\bf d}_{k}o_{s_1,k}{\varphi_{s_1,k}}\right)\left(\sum_{s_2\in\mathcal{S}}{\bar {\bf H}}_{s_2, k}{\bf W}_{s_2, k}{\bf d}_{k}o_{s_2,k}{\varphi_{s_2,k}}\right)^H\right\}\notag\\
        &\textstyle=\sum_{s_1\in\mathcal{S}}\sum_{s_2\in\mathcal{S}}\mathbb{E}\left\{{\bar {\bf H}}_{s_1, k}{\bf W}_{s_1, k}{\bf W}^H_{s_2, k}{\bar {\bf H}}^H_{s_2, k}\varphi_{s_1,k}\varphi^H_{s_2,k}\right\}  o_{s_1,k}o_{s_2,k},\label{Rsig eq}\\
        {\bar {\bf R}}_{{\rm other}, k} &\textstyle=  \mathbb{E}_{{\bf H},{\bf d}, \boldsymbol{\varphi}}\left\{\sum_{j\neq k}\left(\sum_{s_1\in\mathcal{S}}{\bar {\bf H}}_{s_1, k} {\bf W}_{s_1, j}{\bar {\bf d}}^{s_1,k}_{j}o_{s_1,j}{\varphi_{s_1,j,k}}\right)\left(\sum_{s_2\in\mathcal{S}}{\bar {\bf H}}_{s_2, k}{\bf W}_{s_2, j}{\bar {\bf d}}^{s_2,k}_{j}o_{s_2,j}{\varphi_{s_2,j,k}}\right)^H\right\}+ \sigma^2_k{\bf I}\notag\\
&\textstyle\stackrel{(a)}{\approx}\sum_{j\neq k}\sum_{s\in\mathcal{S}}\mathbb{E}\left\{{\bar {\bf H}}_{s, k}{\bf A}_{s} {\bf W}_{s, j}{\bf W}^H_{s, j}{\bf A}^H_{s}{\bar {\bf H}}^H_{s, k}\right\}  o_{s,j}+ \sigma^2_k{\bf I}.\label{Rother eq}
    \end{align}}
\end{figure*}
\vspace{-2mm}
\begin{remark}
From a mathematical perspective, the aforementioned equivalence holds for almost any ${\bar {\bf R}}_{{\rm sig}, k}$ and ${\bar {\bf R}}_{{\rm other}, k}$, and can be regarded as a matrix form of the Quadratic Transform in \cite{shen2018fractional}.
From a communication perspective, $({\text{P3}})$ can be viewed as a generalized WMMSE problem. Therefore, the definition of $ \mathbf{E}_k $ differs from the MSE definition in traditional WMMSE optimization problems, and $ \mathbf{C}_k $ and $ \mathbf{D}_k $ also differ from the MSE weight and MMSE receiver definitions. Moreover, the dimensions of these three terms depend on the covariance decomposition method. Since the former is a generalized form of the latter, when the channel is perfectly known at the satellite side and pre-compensation is perfect, CDWMMSE degenerates to the well-known WMMSE problem \cite{christensen2008weighted, shi2011iteratively}, with $\mathbf{E}_k$, $\mathbf{C}_k$, and $\mathbf{D}_k$ degenerating accordingly.
\end{remark}

\vspace{-3mm}
\subsection{Covariance Decomposition}
\vspace{-1mm}

In problem (P2), the specific expressions for ${\bar {\bf R}}_{{\rm other}, k}\in\mathbb{C}^{N_{\rm R}\times N_{\rm R}}$ and ${\bar {\bf R}}_{{\rm sig}, k}\in\mathbb{C}^{N_{\rm R}\times N_{\rm R}}$ are given in \eqref{Rsig eq} and \eqref{Rother eq}, respectively. 
In \eqref{Rother eq}, term (a) stems from the result $ \mathbb{E}\{{\bar {\bf d}}^{s_1,k}_{j}{\varphi_{s_1,j,k}}({\bar {\bf d}}^{s_2,k}_{j}{\varphi_{s_2,j,k}})^H\} \approx 0 $ derived in our asynchronous interference analysis in \secref{channel sec}. To further simplify these expressions, we can derive
\vspace{-1mm}
\begin{align}
        &\mathbb{E}\left\{{\bar {\bf H}}_{s_1, k}{\bf W}_{s_1, k}{\bf W}^H_{s_2, k}{\bar {\bf H}}^H_{s_2, k}\varphi_{s_1,k}\varphi^H_{s_2,k}\right\}\notag\\
    &=\mathbb{E}\left\{{\bar {\bf u}}_{s_1, k}{\bar {\bf v}}^T_{s_1, k}{\bf W}_{s_1, k}{\bf W}^H_{s_2, k}{\bar {\bf v}}^*_{s_2, k}{\bar {\bf u}}^H_{s_2, k}\varphi_{s_1,k}\varphi^H_{s_2,k}\right\}\notag\\
    & = {\bar {\bf v}}^T_{s_1, k}{\bf W}_{s_1, k}{\bf W}^H_{s_2, k}{\bar {\bf v}}^*_{s_2, k}\mathbb{E}\left\{{\bar {\bf u}}_{s_1, k}{\bar {\bf u}}^H_{s_2, k}\right\}\mathbb{E}\left\{\varphi_{s_1,k}\varphi^H_{s_2,k}\right\}\notag\\
    & =({\bar {\bf v}}^T_{s_1, k}{\bf W}_{s_1, k}{\bf W}^H_{s_2, k}{\bar {\bf v}}^*_{s_2, k}){\boldsymbol{\Delta}}_{s_1,s_2,k},
\end{align}
where ${\bar {\bf v}}^T_{s, k} = {\bf v}^T_{s, k}{\bf F}_s{\bf A}_s\in\mathbb{C}^{1\times B_s}$ and ${\boldsymbol{\Delta}}_{s_1, s_2, k}\in\mathbb{C}^{N_{\rm R}\times N_{\rm R}}$ is given by
\begin{align}
{\boldsymbol{\Delta}}_{s_1, s_2, k} \!=\!\! 
\begin{cases}
    \rho_{s,k}{\bf u}_{s,k}{\bf u}^H_{s,k}+{\tilde  \rho}_{s,k}{\boldsymbol{\Sigma}}_{s,k},\ \ \ s_1\!\!=\!\!s_2\!\!=\!\!s\\
\sqrt{\rho_{s_1,k}\rho_{s_2,k}}{\bar \varphi}_{s_1,k}{\bar \varphi}^H_{s_2,k}{\bf u}_{s_1,k}{\bf u}^H_{s_2,k},   s_1\!\!\neq\!\! s_2,
\end{cases}
\label{eq delta}
\end{align}
where $\rho_{s,k} = \frac{\kappa_{s,k}\gamma_{s,k}}{\kappa_{s,k}+1}$ and ${\tilde \rho}_{s,k} = \frac{\gamma_{s,k}}{\kappa_{s,k}+1}$. 
Based on the above equation, the following simplified expression can be obtained
\begin{align}
    {\bar {\bf R}}_{{\rm sig}, k} &\textstyle= \sum_{s_1\in\mathcal{S}}\sum_{s_2\in\mathcal{S}}({\bf q}^H_{s_1,k,k}{\bf q}_{s_2,k,k}){\boldsymbol{\Delta}}_{s_1, s_2, k},\\
    {\bar {\bf R}}_{{\rm other}, k} &\textstyle= \sum_{j\neq k}\sum_{s\in\mathcal{S}}{\bf q}^H_{s,j,k}{\bf q}_{s,j,k}{\boldsymbol{\Delta}}_{s,s, k} + \sigma^2_k{\bf I},
\end{align}
where ${\bf q}^H_{s,j,k}=o_{s,j}{\bar {\bf v}}^T_{s, k}{\bf W}_{s, j}\in\mathbb{C}^{1\times M_k}$.

To solve problem (P3), deriving a tractable closed-form expression for ${\bar {\bf R}}^{\frac{1}{2}}_{{\rm sig}, k}$ is essential. To this end, we construct the following formula
\begin{lemma}\label{lemma R half}
The matrix ${\bar {\bf R}}_{{\rm sig}, k}$ can be decomposed as ${\bar {\bf R}}_{{\rm sig}, k} = {\bar {\bf R}}^{\frac{1}{2}}_{{\rm sig}, k}({\bar {\bf R}}^{\frac{1}{2}}_{{\rm sig}, k})^H$, where ${\bar {\bf R}}^{\frac{1}{2}}_{{\rm sig}, k}\in\mathbb{C}^{N_{\rm R}\times (SN_{\rm R}+1)M_k}$ is given by
    \vspace{-2mm}
    \begin{align}
        {\bar {\bf R}}^{\frac{1}{2}}_{{\rm sig}, k}
    =\begin{bmatrix}
    \sum_{s=1}^S{\bar \varphi}_{s,k}\sqrt{\rho_{s,k}}{\bf u}_{s,k}{\bf q}^H_{s,k,k} & {\tilde {\boldsymbol{\Sigma}}}_{k}{\bf Q}^H_{k}
    \end{bmatrix},
    \end{align}
    ${\tilde {\boldsymbol{\Sigma}}}_{k} \!=\! \big[ {\tilde {\boldsymbol{\Sigma}}}_{1,k}  \cdots  {\tilde {\boldsymbol{\Sigma}}}_{S,k}\big]$, ${\bf Q}^H_{k} \!=\! {\rm blkdiag}\{{\bf Q}^H_{1,k}, ... ,{\bf Q}^H_{S,k}\}$, and ${\bf Q}^H_{s,k} = {\bf I}_{N_{\rm R}}\otimes{\bf q}^H_{s,k,k}\in\mathbb{C}^{N_{\rm R}\times N_{\rm R}M_k}$.
Here, ${\tilde {\boldsymbol{\Sigma}}}_{s,k}$ is obtained from the following decomposition
\vspace{-2mm}
    \begin{align}
        (1\!-\!{\bar \varphi}_{s,k}{\bar \varphi}^H_{s,k})\rho_{s,k}{\bf u}_{s,k}{\bf u}^H_{s,k}+{\tilde \rho}_{s,k}{\boldsymbol{\Sigma}}_{s,k}={\tilde {\boldsymbol{\Sigma}}}_{s,k}{\tilde {\boldsymbol{\Sigma}}}^H_{s,k},
        \label{eq tilde sigma}
    \end{align}
    which can be achieved by Cholesky decomposition.
\end{lemma}
\begin{pf}
See Appendix \ref{pf R half}.\qedhere
\end{pf}

\vspace{-4.5mm}
\subsection{CDWMMSE Precoding}
\vspace{-1mm}

Based on the decomposition method derived in the previous subsection, in this section, we address the solution to Problem (P3). Due to the nonconvex nature, we adopt an alternating optimization approach, optimizing $\{{\bf D}_k\}_{\forall k}$, $\{{\bf C}_k\}_{\forall k}$, and $\{{\bf W}_{s,k}\}_{\forall s,k}$ while fixing the other two, respectively. Since $\mathbf{D}$ and $\mathbf{C}$ are unconstrained variables, the final expressions are obtained by setting the gradients to zero, as follows:
\vspace{-2mm}
\begin{align}
    &{\bf D}^{\star}_k = ({\bar {\bf R}}_{{\rm sig}, k}+{\bar {\bf R}}_{{\rm other}, k} )^{-1}{\bar {\bf R}}_{{\rm sig}, k}^{\frac{1}{2}},\ \forall k\in\mathcal{K},\label{optimal D eq}\\
    &{\bf C}^{\star}_k = {\bf E}^{-1}_k = {\bf I}+({\bar {\bf R}}_{{\rm sig}, k}^{\frac{1}{2}})^H{\bar {\bf R}}_{{\rm other}, k}^{-1}{\bar {\bf R}}_{{\rm sig}, k}^{\frac{1}{2}},\ \forall k\in\mathcal{K},\label{optimal C eq}
\end{align}
where the expression for ${\bar {\bf R}}_{{\rm sig}, k}^{\frac{1}{2}}$ is given by \lmref{lemma R half}.

With all other variables fixed, we further introduce an $ \eta_k $ for each $ \mathbf{D}_k $ to simplify the solution for the optimal $ \mathbf{W}_{s,k} $, which can be viewed as simultaneously optimizing the scaling of $ \mathbf{W}_{s,k} $ and $ \mathbf{D}_k $. The specific optimization problem is formulated as follows:
\vspace{-1mm}
\begin{align}
({\text P4}):\quad &\min\limits_{\{{\bf W}_{s,k}\}, {\boldsymbol{\eta}}}\ {\sum_{\forall k}}\beta_k\left[{\rm Tr}({\bf C}_k{\bf E}_k)- \log\det({\bf C}_k)\right]\notag\\
&{\rm s.t.}\ \textstyle  \sum_{k\in\mathcal{K}}{\rm Tr}({\bf W}_{s,k}{\bf W}^H_{s,k}) \leq P_s,\ s\in\mathcal{S}.
\label{modified wmmse problem 4}
\end{align}
We expand ${\bar {\bf R}}_{{\rm other}, k}$ as ${\bar {\bf R}}_{{\rm other}, k} = \sum_{j\neq k}{\bar {\bf R}}_{{\rm interf}, j,k} + \sigma^2_k{\bf I}$, 
where ${\bar {\bf R}}_{{\rm interf}, j,k}$ denotes the interference part from UT $ j $ to UT $ k $.
Building upon this, the objective function $ f $ of optimization problem (P4) can be expressed as \eqref{f eq1}. 
\begin{figure*}[b]
    \normalsize
    \vspace{-4mm}
    \hrulefill
    \vspace{-2mm}
\begin{align}
    f &\textstyle= {\sum_{\forall k}}\beta_k{\rm Tr}\left\{{\bf C}_k\left[\frac{1}{\eta_k^2}{\bf D}^H_k\left({\bar {\bf R}}_{{\rm sig}, k}+\sum_{j\neq k}{\bar {\bf R}}_{{\rm interf}, j,k}\right){\bf D}_k - \frac{1}{\eta_k}{\bf D}^H_k{\bar {\bf R}}_{{\rm sig}, k}^{\frac{1}{2}}-\frac{1}{\eta_k}({\bar {\bf R}}_{{\rm sig}, k}^{\frac{1}{2}})^H{\bf D}_k + \frac{\sigma^2_k}{\eta_k^2}{\bf D}^H_k{\bf D}_k\right]\right\},\label{f eq1}\\
    f' &\textstyle={\sum_{\forall k}}\beta_k{\rm Tr}\left\{{\bf C}_k\left[{\bf D}^H_k\left(\frac{{\bar {\bf R}}_{{\rm sig}, k}}{\eta_k^2}+\sum_{j\neq k}\frac{{\bar {\bf R}}_{{\rm interf}, j,k}}{\eta^2_j}\right){\bf D}_k - \frac{1}{\eta_k}{\bf D}^H_k{\bar {\bf R}}_{{\rm sig}, k}^{\frac{1}{2}}-\frac{1}{\eta_k}({\bar {\bf R}}_{{\rm sig}, k}^{\frac{1}{2}})^H{\bf D}_k+ \frac{\sigma^2_k}{\eta_k^2}{\bf D}^H_k{\bf D}_k\right]\right\}.\label{f eq2}
\end{align}
\end{figure*}
Furthermore, we consider approximating the problem to mitigate the obstacles posed by multiple satellite power constraints in its solution. Given that SatCom systems are usually power-limited with relatively minor inter-user interference, we approximate the objective function $ f $ as $ f' $ in \eqref{f eq2}, where the scaling $ \eta_k $ in the interference term for $ \mathbf{D}_k $ is approximated as $ \eta_j $. Additionally, we replace the satellite power constraints with user-specific power constraints, which aligns with our user-centric transmission design in this paper. The combination of these two steps enables the decoupling of precoding designs across UTs, yielding the following optimization problem for UT $k$'s precoding design:
\begin{align}
({\text{P5}}): &\min\limits_{{\bf W}_k, \eta_k}{\rm Tr}\left\{\frac{{\boldsymbol{\Upsilon}}_{k}}{\eta_k^2}\!-\! \frac{\beta_k}{\eta_k}{\bf C}_k\!\left({\bf D}^H_k{\bar {\bf R}}_{{\rm sig}, k}^{\frac{1}{2}}\!+\!({\bar {\bf R}}_{{\rm sig}, k}^{\frac{1}{2}})^H{\bf D}_k\right)\!\right\}\notag\\
&{\rm s.t.}\   {\rm Tr}\left({\bf W}_k{\bf W}^H_k\right) \leq {\tilde P}_k,
\label{modified wmmse problem 5}
\end{align}
where ${\bf W}_k=[{\bf W}^T_{1,k},...,{\bf W}^T_{S,k}]^T$ and ${\boldsymbol{\Upsilon}}_{k} \!\!=\!\! \beta_k{\bf C}_k{\bf D}^H_k({\bar {\bf R}}_{{\rm sig}, k}+\sigma^2{\bf I}){\bf D}_k + \sum_{j\neq k}\beta_j{\bf C}_j{\bf D}^H_j{\bar {\bf R}}_{{\rm interf}, k, j}{\bf D}_j{\bf C}_j$.

To obtain the optimal solution to $({\text{P5}})$, we first derive the following result
\vspace{-2mm}
\begin{lemma}\label{lemma gradient part1}
    The gradient of the primary component ${\rm Tr}\{{\boldsymbol{\Upsilon}}_{k}\}$ in the first term of the objective function in $({\text{P5}})$ with respect to $ \mathbf{W} $ is given as follows:
    \begin{flalign}
        &\textstyle\frac{\partial {\rm Tr}\{{\boldsymbol{\Upsilon}}_{k}\}}{\partial {\bf W}^*_k} \!\!=\!\! \left(\!\beta_k{\breve {\bf V}}^H_k{\boldsymbol{\Psi}}_k{\breve {\bf V}}_k \!\!+\!\! \sum_{j\neq k}\beta_j{\tilde {\bf V}}^H_{j,k}{\tilde {\boldsymbol{\Psi}}}_j{\tilde {\bf V}}_{j,k}\!\right){\bf W}_{k},&
    \end{flalign}
    where
    \begin{align}
    {\breve {\bf V}}_k &= {\rm blkdiag}\{o_{1,k}{\bar {\bf v}}^T_{1,k},...,o_{S,k}{\bar {\bf v}}^T_{S,k}\}\in\mathbb{C}^{S\times {\tilde B}},\\
    {\tilde {\bf V}}_{j,k} &= {\rm blkdiag}\{o_{1,k}{\bar {\bf v}}^T_{1,j},...,o_{S,k}{\bar {\bf v}}^T_{S,j}\}\in\mathbb{C}^{S\times {\tilde B}},\\
    {\boldsymbol{\Psi}}_k &= \begin{bmatrix}
{\psi}_{1,1, k} & {\psi}_{1,2, k} & \cdots & {\psi}_{1,S, k} \\
{\psi}_{2,1, k} & {\psi}_{2,2, k} & \cdots & {\psi}_{2,S, k} \\
\vdots & \vdots & \ddots & \vdots \\
{\psi}_{S,1, k} & {\psi}_{S,2, k} & \cdots & {\psi}_{S,S, k}
\end{bmatrix}\in\mathbb{C}^{S\times S},\\
    {\tilde {\boldsymbol{\Psi}}}_j &= {\rm diag}\{{\psi}_{1,1, j},\cdots,{\psi}_{S,S, j}\}\in\mathbb{C}^{S\times S},
\end{align}
in which ${\psi}_{s_2, s_1,k} \!=\! {\rm Tr}\left({\bf D}_k{\bf C}_k{\bf D}^H_k{\boldsymbol{\Delta}}_{s_1, s_2, k}\right)$ and ${\tilde B} \!= \!\sum_{s=1}^S B_s$.
\end{lemma}
\begin{pf}
See Appendix \ref{pf gradient part1}.
\end{pf}

\begin{lemma}\label{lemma gradient part2}
    The gradient of the primary component in the second term of the objective function in $({\text{P5}})$ with respect to $\mathbf{W}$ is given as follows:
    \begin{align}
    &{\partial {\rm Tr}({\bf C}_k({\bar {\bf R}}_{{\rm sig}, k}^{\frac{1}{2}})^H{\bf D}_k]))}/{\partial {\bf W}^*_{k}} = {\breve {\bf V}}^H_k{\bf T}_{k},
    \end{align}
where ${\bf T}_{k}$ is given by
    \begin{align}
    &{\bf T}_{k} = 
    \begin{bmatrix}
    {\bf t}_{1,k}&
    \cdots&
    {\bf t}_{S,k}
\end{bmatrix}^T\in\mathbb{C}^{S\times M_k},\\
    &\textstyle{\bf t}^T_{s,k} ={\bar \varphi}^H_{s,k}\sqrt{\rho_{s,k}}{\bf u}^H_{s,k}{\bf D}_k{\bar {\bf C}}_k+\sum_{n=1}^{N_{\rm R}}
    {\boldsymbol{\sigma}}^H_{s,k,n}{\bf D}_k{\tilde {\bf C}}_{s,k,n}.
    \label{eq t}
    \end{align}
    The vector ${\boldsymbol{\sigma}}^H_{s,k,n}$ and matrices ${\bar {\bf C}}_k\in\mathbb{C}^{L_k\times M_k}$, ${\tilde {\bf C}}_{s,k,n}\in\mathbb{C}^{L_k\times M_k}$ are constructed via the following expressions
    \begin{align}
        &{\tilde {\boldsymbol{\Sigma}}}^H_{s,k}= \big[
       {\boldsymbol{\sigma}}_{s,k,1} 
       \cdots
       {\boldsymbol{\sigma}}_{s,k,N_{\rm R}}
   \big]^H, {\bf C}_{k} = 
   \begin{bmatrix}
      {\bar {\bf C}}_k & {\tilde {\bf C}}_{1,k} & \cdots & {\tilde {\bf C}}_{S,k}
   \end{bmatrix}, \nonumber \\
   &\qquad{\tilde {\bf C}}_{s,k} = 
   \begin{bmatrix}
       {\tilde {\bf C}}_{s,k, 1} & \cdots & {\tilde {\bf C}}_{s,k, N_{\rm R}}
   \end{bmatrix}\in\mathbb{C}^{L_k\times N_{\rm R}M_k}. \nonumber 
    \end{align}
    \end{lemma}
\begin{pf}
See Appendix \ref{pf gradient part2}.
\end{pf}

Based on the aforementioned lemmas, we can obtain the following form of the optimal solution
\begin{ppn}\label{ppn optimal solution}
    The following solution achieves the optimality for optimization problem $({\text{P5}})$
    \begin{align}
        &{\bf W}^{\rm CDWM}_{k} = \eta^{\star}_k{\bar {\bf W}}_k,\ {\bar {\bf W}}_k=
        ({\boldsymbol{\Xi}}_k+ {\tilde \beta}_k{\bf I})^{-1}{\breve {\bf V}}^H_k{\bf T}_{k},\label{closed form}\\
        &{\tilde \beta}_k=\frac{\beta_k\sigma_k^2}{{\tilde P}_k}{\rm Tr}\{{\bf D}_k{\bf C}_k{\bf D}^H_k\},\ \eta^{\star}_k = {\tilde P}_k/{\|{\bar {\bf W}_k}\|^2_F},\\
        &\textstyle{\boldsymbol{\Xi}}_k = \beta_k{\breve {\bf V}}^H_k{\boldsymbol{\Psi}}_k{\breve {\bf V}}_k \!+\! \sum_{j\neq k}\beta_j{\tilde {\bf V}}^H_{j,k}{\tilde {\boldsymbol{\Psi}}}_j{\tilde {\bf V}}_{j,k}.
    \end{align}
\end{ppn}
\begin{pf}
See Appendix \ref{pf optimal solution}.
\end{pf}
\vspace{-2mm}

Combining the optimal expressions in \eqref{optimal D eq}, \eqref{optimal C eq}, and \eqref{closed form}, we can formulate the MSMS precoding algorithm as \algref{MS-JoCDWM Algorithm}, where $\text{MS}^2$ denotes MSMS. 
In this algorithm, the dominant computational burden comes from Steps \ref{alg C eq}, \ref{alg Xi eq}, and \ref{alg W eq}, which together determine the overall computational complexity on the order of $\mathcal{O}(I_{\rm max}(KS^3N_{\rm R}^3M^3 + K^2{\tilde B}^2S+K{\tilde B}^3))$, where $M$ denotes the average number of data streams per UT. With the architecture of beamspace MIMO transmission, this complexity is independent of the number of transmit antennas $N_{\rm T}$ and depends only on the number of active beams ${\tilde B} = \sum_{s=1}^S B_s$, highlighting the computational efficiency. Specifically, since $N_{\rm T} \gg K \gg N_{\rm R}$, the computational complexity of beamspace MIMO transmission relative to conventional MIMO scales approximately as $(\frac{{\tilde B}}{SN_{\rm T}})^3$. For instance, this ratio is about $0.7\%$ with $B_s=48$ and $N_{\rm T}=256$.

{
\setlength{\textfloatsep}{-3pt}
\setlength{\intextsep}{-3pt}
\begin{algorithm}[!t]
    \caption{$\text{MS}^2$CDWM Precoding Algorithm}
    \label{MS-JoCDWM Algorithm}
    \begin{spacing}{1.2}
    \KwIn{${\mathcal H}_{\rm MS}=\{\gamma_{s,k},\kappa_{s,k},{\boldsymbol{\theta}}_{s,k},{\boldsymbol{\Sigma}}_{s,k},{\bar \varphi}_{s,k}\}$, $\{P_s\}$, $\{\sigma^2_k, \beta_k\}$, $\{{\bf A}_{s}\}$, $\{{\bf F}_{s}\}$,  $\{{\bf O}_{s}\}$, $I_{\rm max}$, $\chi$} 
    \KwOut{$\{{\bf W}_{s,k}\}_{\forall s,k}$}
    Construct $\{{\bf v}_{s,k}\}$ and $\{{\bf u}_{s,k}\}$ using \eqref{eq v} and \eqref{eq u}\;
    ${\bar {\bf v}}^T_{s, k} = {\bf v}^T_{s, k}{\bf F}_s{\bf A}_s,\ \forall s,k$\;
    ${\breve {\bf V}}_k = {\rm blkdiag}\{o_{1,k}{\bar {\bf v}}^T_{1,k},...,o_{S,k}{\bar {\bf v}}^T_{S,k}\},\ \forall k$\;
    ${\tilde {\bf V}}_{j,k} = {\rm blkdiag}\{o_{1,k}{\bar {\bf v}}^T_{1,j},...,o_{S,k}{\bar {\bf v}}^T_{S,j}\},\ \forall j,k$\;
    Compute $\{{\boldsymbol{\Delta}}_{s_1, s_2, k}\}$ and $\{{\tilde {\boldsymbol{\Sigma}}}_{s,k}\}$ using \eqref{eq delta} and \eqref{eq tilde sigma}\;
    Initialize $\{{\bf W}_k\}$, $n=0$\;
    Initialize ${\tilde P}_k = (\sum_{s\in{\mathcal S}}P_s)/K$, ${\bf E}_k = \chi{\bf I}$, $\forall k$\;
    \Repeat{$n\geq I_{\rm max}$ \text{or} $\sum_{k=1}^K\beta_k\log_2\left[\frac{\det({\bf E}'_k)}{\det({\bf E}_k)}\right]<\epsilon$}{$n = n+1$\;
    ${\bf E}'_k = {\bf E}_k,\ \forall k$\;
    $    {\bf q}^H_{s,k,k}= o_{s,k}({\bf v}_{s,k}^{\rm sat})^T{\bf F}_s{\bf A}_{s} {\bf W}_{s, k},\ \forall s,k$\;
    $
    {\bf q}^H_{s,j,k}=o_{s,j}({\bf v}_{s,k}^{\rm sat})^T{\bf F}_s{\bf A}_{s} {\bf W}_{s, j},\ \forall s,j,k$\;
    ${\bar {\bf R}}_{{\rm other}, k} = \sum\limits_{s\in\mathcal{S}}\sum\limits_{j\neq k}{\bf q}^H_{s,j,k}{\bf q}_{s,j,k}{\boldsymbol{\Delta}}_{s,s, k} + \sigma^2_k{\bf I},\ \forall k$\;
    ${\bar {\bf R}}_{{\rm sig}, k} = \sum\limits_{s_1\in\mathcal{S}}\sum\limits_{s_2\in\mathcal{S}}({\bf q}^H_{s_1,k,k}{\bf q}_{s_2,k,k}){\boldsymbol{\Delta}}_{s_1, s_2, k},\ \forall k$\;
    ${\bar {\bf R}}^{\frac{1}{2}}_{{\rm sig}, k}
    =\begin{bmatrix}
    \sum\limits_{s=1}^S{\bar \varphi}_{s,k}\sqrt{\rho_{s,k}}{\bf u}_{s,k}{\bf q}^H_{s,k,k} & {\tilde {\boldsymbol{\Sigma}}}_{k}{\bf Q}^H_{k}
    \end{bmatrix},\ \forall k$\;
    ${\bf D}_k = ({\bar {\bf R}}_{{\rm sig}, k}+{\bar {\bf R}}_{{\rm other}, k} )^{-1}{\bar {\bf R}}_{{\rm sig}, k}^{\frac{1}{2}},\ \forall k$\;
    Compute ${\bf E}_k$ using \eqref{MSE loss function},\ $\forall k$\;
    ${\bf C}_k = {\bf E}^{-1}_k,\ \forall k$\label{alg C eq}\;
    ${\psi}_{s_2, s_1,k} = {\rm Tr}\left({\bf D}_k{\bf C}_k{\bf D}^H_k{\boldsymbol{\Delta}}_{s_1, s_2, k}\right),\ \forall s_1,s_2,k$\;
    ${\boldsymbol{\Xi}}_k = \beta_k{\breve {\bf V}}^H_k{\boldsymbol{\Psi}}_k{\breve {\bf V}}_k + \sum\limits_{j\neq k}\beta_j{\tilde {\bf V}}^H_{j,k}{\tilde {\boldsymbol{\Psi}}}_j{\tilde {\bf V}}_{j,k},\ \forall k$\label{alg Xi eq}\;
    ${\tilde \beta}_k=\frac{\beta_k\sigma_k^2}{{\tilde P}_k}{\rm Tr}\{{\bf D}_k{\bf C}_k{\bf D}^H_k\},\ \forall k$\;
    Compute ${\bf t}_{s,k}$ using \eqref{eq t}, $\forall s,k$\;
    ${\bar {\bf W}}_{k} =\eta_k\left({\boldsymbol{\Xi}}_k+ {\tilde \beta}_k{\bf I}\right)^{-1}{\breve {\bf V}}^H_k{\bf T}_{k},\ \forall k$.\label{A1 closed form step}\label{alg W eq}\;
    $\eta_k = \sqrt{\frac{{\tilde P}_k}{\|{\bar {\bf W}}_{k}\|^2_F}},\ {\bf W}'_{k}=\eta_k{\bar {\bf W}}_{k},\ \forall k$\;
    }
    ${\bf W}_{s,k} \!=\! \min\!\left(\!\sqrt{\frac{P_s}{\sum_{k\in\mathcal{K}}\|{\bf W}'_{s,k}\|^2_F}}, 1\!\right){\bf W}'_{s,k}$ $\forall s,k$\;\label{sat power step}
    \end{spacing}
\end{algorithm}
}

\vspace{-1mm}
\section{Multi-Satellite Multi-Stream  \\Closed-Form Precoding Design}\label{CF precoding sec}
\vspace{-1mm}
The above iterative algorithms improve performance through multiple iterations but at the cost of reduced computational efficiency. To address this, we propose heuristic closed-form schemes in this section to achieve an effective tradeoff between computational efficiency and performance.

\vspace{-3mm}
\subsection{$\text{MS}^2$CDM Precoding}
\label{CDM sec}
\vspace{-1mm}
Observing the iterative process of \algref{MS-JoCDWM Algorithm} and closed-form \eqref{closed form}, we deduce that the iterations of \algref{MS-JoCDWM Algorithm} aim to solve for variables $\{{\bf C}_k\}$ and $\{{\bf D}_k\}$. Inspired by this, we directly assign their values to avoid iteration. Specifically, we first consider the case where $M_k=N_{\rm R},\ \forall k$, assigning $\{{\bf C}_k\}$ and $\{{\bf D}_k\}$ as ${\bf D}_{k[:,1:N_{\rm R}]} = {\bf C}_{k[1:N_{\rm R},1:N_{\rm R}]} = {\bf I}_{N_{\rm R}}$, 
where all unassigned elements are set to zero. This yields the following closed-form precoding 
\begin{align}
    &{\bf W}^{\rm CDM}_k\!\!=\!{\tilde \eta}_{k}{\tilde {\bf W}}_k,\ 
    {\tilde {\bf W}}_k = \left({\tilde {\boldsymbol{\Xi}}}_k+ {\breve \beta}_k{\bf I}\right)^{-1}{\breve {\bf V}}^H_k{\breve {\bf T}}_{k},\\
    &\quad\textstyle{\tilde {\boldsymbol{\Xi}}}_k \!=\! \beta_k{\breve {\bf V}}^H_k{\breve {\boldsymbol{\Delta}}}_k{\breve {\bf V}}_k \!+\! \beta_j\sum_{j\neq k}{\tilde {\bf V}}^H_{j,k}{\tilde {\boldsymbol{\Delta}}}_j{\tilde {\bf V}}_{j,k},
\end{align}
where ${\tilde \eta}_{k} = \sqrt{{{\tilde P}_k}/{\|{\tilde {\bf W}_k}\|^2_F}}$, $ {\breve{\delta}}_{s_2, s_1,k} = {\rm Tr}\left({\boldsymbol{\Delta}}_{s_1, s_2, k}\right)$, $\ {\breve \beta}_k = {\beta_k\sigma_k^2N_{\rm R}}/{\tilde P}_k$, ${\tilde {\boldsymbol{\Delta}}}_j = {\rm diag}\{{\breve{\delta}}_{1,1, j},\cdots,{\breve{\delta}}_{S,S, j}\}\in\mathbb{C}^{S\times S}$, ${\breve {\bf T}}_{k} = 
\big[{\breve {\bf t}}_{1,k},
\cdots,
{\breve {\bf t}}_{S,k}
\big]^T$, ${\breve {\bf t}}^T_{s,k} ={\bar \varphi}^H_{s,k}\sqrt{\frac{\kappa_{s,k}\gamma_{s,k}}{\kappa_{s,k}+1}}{\bf u}^H_{s,k}$ and
\begin{align}
{\breve {\boldsymbol{\Delta}}}_k = \begin{bmatrix}
\breve{\delta}_{1,1, k} & \breve{\delta}_{1,2, k} & \cdots & \breve{\delta}_{1,S, k} \\
\vdots & \vdots & \ddots & \vdots \\
{\breve{\delta}}_{S,1, k} & {\breve{\delta}}_{S,2, k} & \cdots & {\breve{\delta}}_{S,S, k}
\end{bmatrix}\in\mathbb{C}^{S\times S}.
\end{align}

When $M_k < N_{\rm R}$, we activate only the first $M_k$ stream precoders, enabling the precoder to adapt to the number of streams
\begin{align}
    \textstyle{\bf W}^{\rm CDM}_k\!\!=\!{\tilde \eta}_{k}{\tilde {\bf W}}_{k[:,\ 1:M_k]},\ {\tilde \eta}_{k} = \sqrt{{{\tilde P}_k}/{\|{\tilde {\bf W}}_{k[:,\ 1:M_k]}\|^2_F}}.
\end{align}
The closed-form precoding can be adjusted to satisfy the satellite power constraints using the approach in Step \ref{sat power step} of \algref{MS-JoCDWM Algorithm}.
We refer to the precoding method as $\text{MS}^2$CDM, as its relationship to $\text{MS}^2$CDWM is similar to that between MMSE and WMMSE.
The computational complexity of this method is $\mathcal{O}(K^2{\tilde B}^2S+K{\tilde B}^3)$, which is significantly lower than that of $\text{MS}^2$CDWM and does not require any iterations.

\vspace{-3mm}
\subsection{Location Information-Based Precoding}
\label{LIB sec}
\vspace{-1mm}

For satellites, if the UT location can be obtained, for example, via the GNSS, the relative position between the satellite and the UT can be calculated in conjunction with the ephemeris. On this basis, by measuring its own attitude, the satellite can acquire the angle information $\{\theta^{\rm t}_{s,k},\phi^{\rm t}_{s,k}\}$. For a specified UT, based on the transmit steering vectors $\{{\bf v}_{s,k}\}$ formed by these angles, each satellite can design precoding for single-stream data transmission to maximize the received power. Considering that the multi-path channel composed of multi-satellite channels needs to support multi-stream transmission for the UT, we distribute the multiple streams required by the UT across these satellites. Specifically, the algorithm for this precoding is detailed in \algref{LI Precoding Algorithm}.
The computational complexity of this algorithm is $\mathcal{O}(K\sum_{s=1}^{S}Q_sN_{\rm T})$.

\begin{algorithm}[!t]
    \caption{LIB Precoding Algorithm}
    \label{LI Precoding Algorithm}
    \begin{spacing}{0.8}
    \KwIn{$\{\theta^{\rm t}_{s,k},\phi^{\rm t}_{s,k}\}$, $\{{\bf A}_{s},{\bf F}_{s},{\bf O}_{s},P_s\}$} 
    \KwOut{$\{{\bf W}_{s,k}\}_{\forall s,k}$}
    Construct $\{{\bf v}_{s,k}\}$ using \eqref{eq v}\;
    Initialize ${\bf W}_{s,k} = {\bf 0}_{B_s\times M_k,\ \forall s,k}$\;
    \For{$k\in \mathcal{K}$}{
    $b=1$\;
    \For{$s\in \mathcal{S}$}{
        \If{$o_{s,k}=1$}{
            ${\bf W}'_{s,k[:,\ b]} = {\bf A}^H_s{\bf F}^H_s{\bf v}^{*}_{s, k}$\;
            \If{$b=M_k$}{
                $b = 1$\;
            }
            \Else{$b = b+1$\;}
        }
    }
    }
    ${\bf W}_{s,k} = \sqrt{\frac{P_s}{\sum_{k\in\mathcal{K}}\|{\bf W}'_{s,k}\|^2_F}}\cdot{\bf W}'_{s,k}$ $\forall s,k$\;\label{sat power step}
    \end{spacing}
\end{algorithm}

\vspace{-2mm}
\section{Statistical CSI-Based Satellite Clustering
\\and Beam Selection}\label{scheduling sec}


In the previous section, we developed the precoding design method under given satellite clustering and beam selection. With the aforementioned precoding design, solving for the optimal satellite clustering and beam selection in $({\text P1})$ remains an NP-hard mixed-integer non-convex optimization problem. Considering that the precoding design involves iterative procedures and the dimensions of the integer variables $\{{\bf O}_s\}$ and $\{{\bf A}_s\}$ are large, employing common solution methods—such as branch-and-bound or optimal exhaustive search—results in unacceptably high computational complexity. Moreover, because precoding can effectively mitigate interference and satellite links are power limited due to severe path loss, the power of the beam-domain equivalent channel becomes particularly important. Accordingly, by exploiting inherent multi-satellite channel characteristics and focusing on channel power enhancement, we design efficient user-centric satellite clustering and beam selection schemes in this section.

\vspace{-3mm}
\subsection{User-Centric Satellite Clustering}
\vspace{-1mm}
\label{clustering}
We adopt the access point selection algorithm from \cite{9174860} as the core of the satellite clustering method, which computes the optimized variable $\{{\bf O}_{s}\}$. This approach relies solely on large-scale parameters ${\gamma_{s,k}}$ from sCSI and prioritizes power allocation, making it inherently suitable for power-constrained SatComs based on a sCSI framework. However, as it does not fully align with our considered scenario, we introduce corresponding enhancements. To ensure each UT receives service from a specified number of satellites, we modify the algorithm to set the minimum number of serving satellites per UT to $S_k$, establishing a lower bound. Additionally, to impose an upper bound, we design a mechanism to remove excess serving satellites for each UT. Specifically, the algorithm incorporates the following key components:
\begin{itemize}
    \item \textbf{Main Procedure:} Iterate over all UTs. For each UT, sequentially select satellites in descending order of average channel power. If the selected satellite has not reached its maximum serving UT capacity, directly establish the association. Otherwise, compete with other UTs served by that satellite. If the number of competition failures (either initiating or receiving) reaches a predefined threshold, forcibly associate with the remaining unselected satellites.
    \item \textbf{Competition:} Except for forcibly established associations, the UT compares channel power with other UTs associated with the selected satellite and eliminates the one with the poorest channel power.
    \item \textbf{Redundant Satellite Removal:} Competition outcomes may lead to certain UTs monopolizing an excessive number of satellites. To address this, after the main loop, we iterate through each UT, sort satellites by power, and remove associations with the least favorable satellites until the number of satellites per UT meets the requirement.
\end{itemize}
Due to space limitations, the procedure of this algorithm is omitted here.

\vspace{-2mm}
\subsection{Low-Complexity Beam Selection}
\vspace{-1mm}

\begin{algorithm}[!t]
    \caption{Two-Stage Beam Selection Algorithm}
    \label{Beam Selection Algorithm}
    \begin{spacing}{0.8}
    \KwIn{$\{\gamma_{s,k}, {\theta^{\rm t}_{s,k}},{\phi^{\rm t}_{s,k}}\}$, $\{{\bf F}_{s},{\bf O}_{s}\}$} 
    \KwOut{$\{{\bf A}_{s}\}$}
    Construct $\{{\bf v}_{s,k}\}$ using \eqref{eq v}\;
    Initialize ${\bf A}_{s} = {\bf 0}_{Q_s\times B_s,\ \forall s}$\;
    \For{$s\in \mathcal{S}$}{
    $b = 1$\;
    $\mathcal{Q}_s=\{1,...,Q_s\}$\;
    $\mathcal{K}'_s=\mathcal{K}_s$\;
    \While{$\mathcal{K}'_s\neq \emptyset$}{
    $k^{\star} = \arg\max_{k\in\mathcal{K}'_s} \gamma_{s,k}$\;
    $q^\star = \arg\max\limits_{b\in\mathcal{Q}_s}\left|\frac{{\rm sinc}\left(N_{\rm TV} \vartheta^b_{s,k^{\star}}\right)}{{\rm sinc}\left(\vartheta^b_{s,k^{\star}}\right)}\right|\!\cdot\!\left|\frac{{\rm sinc}\left(N_{\rm TH} {\tilde \vartheta}^b_{s,k^{\star}}\right)}{{\rm sinc}\left({\tilde \vartheta}^b_{s,k^{\star}}\right)}\right|$\;
    ${\bf A}_{s[q^\star,b]} = 1$\;
    $b = b+1$\;
    ${\mathcal{Q}}_s = {\mathcal{Q}}_s\setminus \{q^\star\}$\;
    $\mathcal{K}'_s = \mathcal{K}'_s \setminus \{k^{\star}\}$\;
    }
    \Repeat{$b=B_s$}{
        $q^\star =$\\ $ \arg\max\limits_{b\in\mathcal{Q}_s}\sum\limits_{k\in\mathcal{K}_s}\!\!\!\gamma_{s,k}\!\left|\frac{{\rm sinc}\left(N_{\rm TV} \vartheta^b_{s,k}\right)}{{\rm sinc}\left(\vartheta^b_{s,k}\right)}\right|\!\cdot\!\left|\frac{{\rm sinc}\left(N_{\rm TH} {\tilde \vartheta}^b_{s,k}\right)}{{\rm sinc}\left({\tilde \vartheta}^b_{s,k}\right)}\right|$\;
        ${\bf A}_{s[q^\star,b]} = 1$\;
        $b = b+1$\;
        ${\mathcal{Q}}_s = {\mathcal{Q}}_s\setminus \{q^\star\}$\;
    }
    }
    \end{spacing}
\end{algorithm}
\vspace{-1mm}
\begin{table}[!t]
    \centering
    \caption{Simulation Parameters \cite{3gpp_tr_38_821, 3GPP_TR_38_811, SpaceX_Gen2_2021, 9998075, 9815679, 10440321}}
    \vspace{-2mm}
     \resizebox{0.48\textwidth}{!}{%
    \begin{tabular}{ll}  
    \toprule
    \textbf{Parameter} & \textbf{Value} \\ 
    \midrule
    Carrier frequency & 2 GHz \\ 
    Subcarrier spacing & 30kHz\\
    UT noise figure & 7 dB \\ 
    UT antenna temperature & 290 K \\ 
    Coverage radius & 800 km \\ 
    Number of UTs $K$ & 12 - 48 (default: 48) \\ 
    Maximum number of served UTs $K^{\rm max}_s$& 36 - 48 (36) \\
    Number of cooperating satellites $S$ & 5\\
    Number of serving satellites $S_k$ & 1 - 5 (3)\\
    Number of beams activated per satellite & $K_s$ - 256 (48)\\
    Distribution of UTs & Uniform \\
    \midrule
    Per-element gain of TX antennas & 6dBi \\ 
    Gain of RX antennas & 0dBi \\
    Transmit antenna size & $N_{\rm TV}=N_{\rm TH}=16$\\
    UT antenna size & $N_{\rm RV}\!=\!N_{\rm RH}\in\{1, 2, 3, 4\}$ (2)\\
    Variance of phase error $\zeta^2_{s,k}$ & 0.5\\
    \midrule
        Satellites altitude & 600 km \\ 
    Constellation Type & Walker-Delta \\ 
    Orbital Planes & 28 \\ 
    Satellites Per Plane & 60 \\ 
    Inclination (degrees) & 53 \\ 
    \bottomrule
    \end{tabular}}
    \label{Simulation Parameter Table}
    \vspace{-3mm}
\end{table}

\subsubsection{Algorithm Overview}From the precoding perspective, beam selection essentially constructs a dimension-reduced equivalent channel. Analyzing the properties of satellite channels leads to the design method for beam selection in SatComs. Satellite channels are dominated by the LoS component, while the large path loss emphasizes the importance of channel power. Based on this, we propose a two-stage beam selection algorithm. In the first stage, we determine UT priorities according to channel power and allocate the strongest beam to each UT sequentially. In the second stage, from the remaining beams, we select the one that maximizes the enhancement of the equivalent channel power for all UTs.

\subsubsection{Low-Complexity Design}Leveraging the LoS path-dominant nature of satellite channels, we derive the following simplified computation for the equivalent channel elements to significantly reduce the beam-selection complexity.
\begin{align}
    \begin{split}
    &\|{\bf H}_{s,k}{\bf f}_b\|^2_2= {\rm Tr}\{{\bf H}_{s,k}{\bf f}_b{\bf f}^H_b{\bf H}^H_{s,k}\} 
    \\&= {\rm Tr}\{{\bf u}_{s,k}{\bf v}^T_{s,k}{\bf f}_b{\bf f}^H_b{\bf v}^{*}_{s,k}{\bf u}^H_{s,k}\}= \gamma_{s,k}{\bf v}^T_{s,k}{\bf f}_b{\bf f}^H_b{\bf v}^{*}_{s,k}\\
    & = \gamma_{s,k} \left|\frac{{\rm sinc}(N_{\rm TV} \vartheta^b_{s,k})}{{\rm sinc}(\vartheta^b_{s,k})}\right|\times \left|\frac{{\rm sinc}(N_{\rm TH} {\tilde \vartheta}^b_{s,k})}{{\rm sinc}({\tilde \vartheta}^b_{s,k})}\right|,
    \end{split}
\end{align}
where ${\rm sinc}(x) = \frac{\sin(\pi x)}{\pi x}$, $\vartheta^b_{s,k} = [\cos ({\theta^{\rm t}_{s,k}})-\varpi_{s,b}]/2$, and ${\tilde \vartheta}^b_{s,k} =[\sin(\smash{\theta^{\rm t}_{s,k}})\cos(\phi^{\rm t}_{s,k})-{\tilde \varpi}_{s,b}]/2$.
In summary, the algorithm for beam selection is detailed in \algref{Beam Selection Algorithm}, and its computational complexity is $\mathcal{O}(K\sum_{s=1}^{S}Q_s)$. 
As such algorithm requires only sCSI like position information and large-scale fading parameters, the selected beams can be applicable across multiple subcarriers.
Notably, the algorithm can be executed at the master satellite and disseminated to the secondary satellites, or computed locally at each satellite and aggregated at the master satellite via ISLs. Its reliance on sCSI makes it relatively robust to ISL delays.

\vspace{-2mm}
\section{Simulation Results}
\label{result sec}
\vspace{-1mm}

We use the QuaDRiGa channel simulator to generate the geographical locations of UTs and satellites, their transmit and receive antennas and orientations, the propagation environment, and finally the channels \cite{6902008,9815679,QuaDRiGa2023}. This channel simulator is recognized in 3GPP channel modeling specifications \cite{3GPP_TR_38_811}, and the `QuaDRiGa\_NTN\_Urban\_LOS' scenario adopted here models the LoS and NLoS powers using a Rician distribution, which is consistent with \eqref{frequency channel model eq} and the 3GPP channel model. The simulator is adapted to achieve closer alignment with the standard channel model and parameters and to better highlight the conclusions of this study. For example, we prevent rapid fluctuations of the K-factor over short time intervals. In channel model \eqref{frequency channel model eq}, the sCSI parameter $\{\boldsymbol{\theta}_{s,k}\}$ at the base station is estimated from transceiver positions and orientations, while $\{\gamma_{s,k}\}$ and $\{{\boldsymbol{\Sigma}}_{s,k}\}$ are estimated from channel samples.
Simulation parameters are detailed in Table \ref{Simulation Parameter Table}. The cooperative transmission coverage radius is configured to 800 km, which does not alter the performance outcomes of the proposed method. Monte Carlo simulations are performed, with each trial randomly selecting a central point within the constellation’s coverage area. A circular region is defined with the specified radius, and the $S$ nearest satellites to this center are chosen for cooperative transmission. We assume uniform subcarrier power constraints across all satellites, denoted as $P_s = P_{\rm TX}, \forall s$, with $P_{\rm TX}$ ranging from 20 dBm to 50 dBm, accommodating various link budgets arising from different transceiver setups \cite{3gpp_tr_38_821}. Performance is evaluated using the average sum rate $R_{\rm E} = \sum_{k\in\mathcal{K}}\mathbb{E}\{R_k\}$.
Without loss of generality, we model the phase error in \eqref{received signal eq1} as $\varphi_{s,k} = {\rm e}^{j\varrho_{s,k}}$ and $\varrho_{s,k} \sim \mathcal{N}(0, \zeta^2_{s,k})$ \cite{10596023, wang2021resource}.

This section compares the following schemes:
\begin{itemize}
\item `\textbf{$\text{MS}^2$DFT}': Each serving satellite uses one DFT beam per UT to transmit one stream. Some satellites diliver the same stream if $S_k>M_k$. 
\item `\textbf{LIB}': 
The location information-based precoding method proposed in \secref{LIB sec}. 
\item `\textbf{$\text{MS}^2$CDM}': The heuristic closed-form CDMMSE precoding method proposed in \secref{CDM sec}.
\item `\textbf{$\text{MS}^2$CDWM}': The proposed iterative precoding method in \algref{MS-JoCDWM Algorithm} of \secref{CDWM sec}.
\item `\textbf{LCMS}', `\textbf{NearOpt}\cite{gao2016near}', and `\textbf{Full}':
The beam selection scheme proposed in \algref{Beam Selection Algorithm}, the well-known near-optimal beam selection scheme in \cite{gao2016near}, and the full-beam activation scheme (with performance equivalent to conventional MIMO transmission).
\end{itemize}
All schemes employ the user-centric clustering method in \secref{clustering}. If the beam selection algorithm is not explicitly specified, the proposed LCMS algorithm is used.


\begin{figure}[!t]  
    \centering
    \captionsetup{font=footnotesize}
    \includegraphics[width=0.7\linewidth]{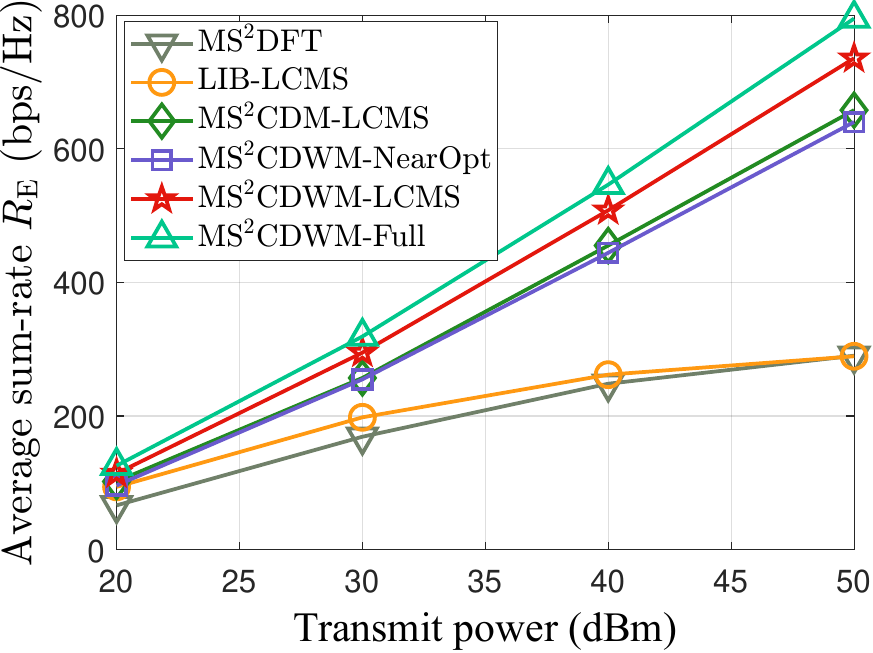}
    \caption{Average sum rate vs $P_{\rm T}$, $M_k=2$, $S_k=3$.}  
    \label{fig SNR S}  
    \vspace{-5mm}
\end{figure}
\begin{figure}[!t]  
    \centering
    \captionsetup{font=footnotesize}
\includegraphics[width=0.7\linewidth]{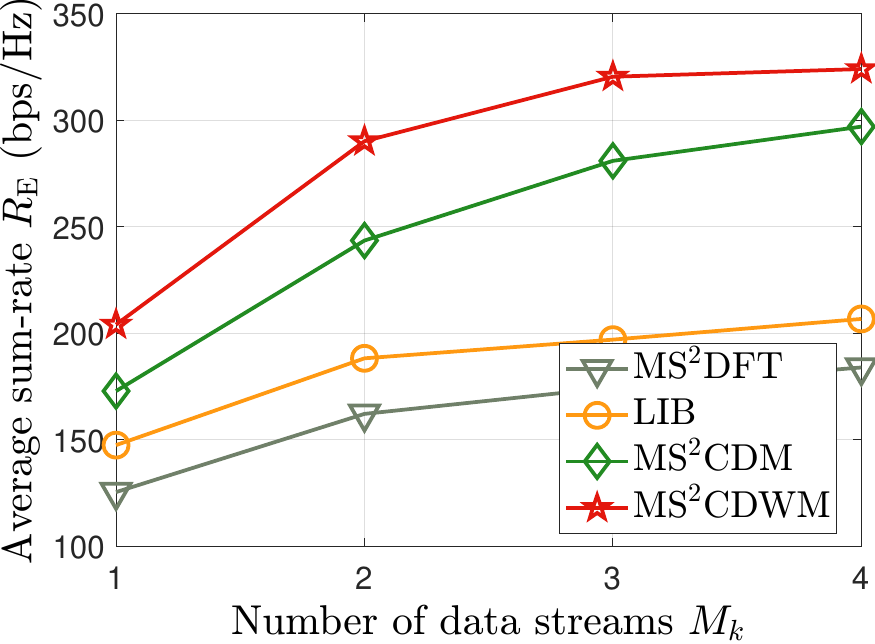}
    \caption{Average sum rate vs $M_k$, $S_k=4$, $P_{
    \rm T}=30{\rm dBm}$.}  
    \label{fig M}  
    \vspace{-6mm}
\end{figure}


\figref{fig SNR S} illustrates the transmission performance variation with transmit power for a given number of data streams ($M_k = 2$), when each UT is served by either two or three satellites. The performance of all methods improves with increasing power, with the CDWM and CDM series algorithms exhibiting stronger gains due to their effective interference suppression. Notably, comparing $\text{MS}^2$DFT, $\text{MS}^2$CDWM-LCMS, and $\text{MS}^2$CDWM-Full reveals that beamspace transmission methods significantly outperform conventional earth-moving beamforming, while achieving comparable performance to antenna-domain precoding with reduced computational complexity and deployment costs. This favorable tradeoff stems from the strong sparsity in the beamspace, driven by the LoS path dominance of satellite channels. \figref{fig M} shows the impact of the number of data streams on performance under sufficient cooperative satellite and receive antenna counts. Increasing the number of data streams can improve the sum rate, but the gain is limited by the receiver capability, such as the number of receive antennas.

\begin{figure}[!t]  
    \centering
    \captionsetup{font=footnotesize}
    \begin{subfigure}{0.24\textwidth}  
        \centering
        \includegraphics[width=\linewidth]{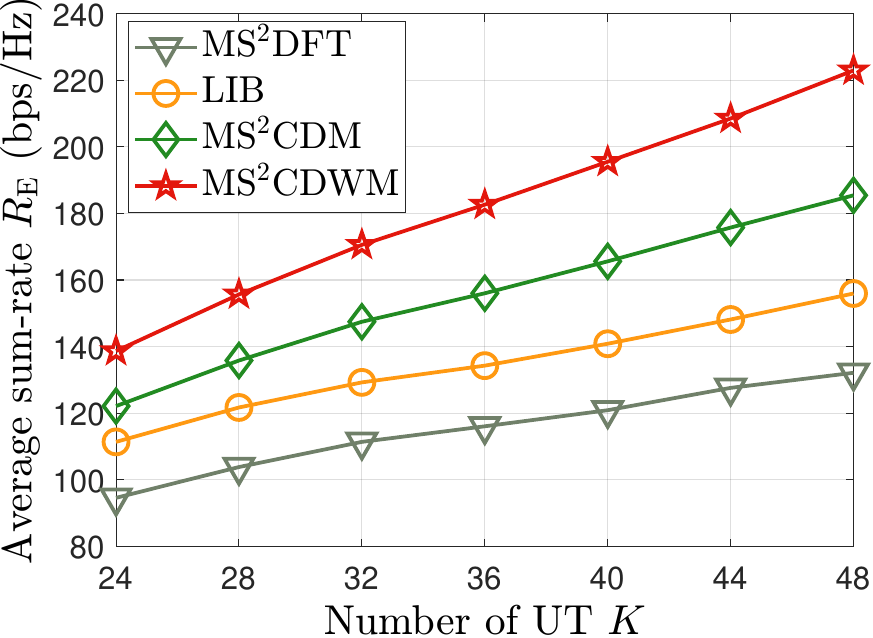}  
        \caption{$M_k=1$.}  
    \end{subfigure}
    \hfill  
    \begin{subfigure}{0.24\textwidth}
        \centering
\includegraphics[width=\linewidth]{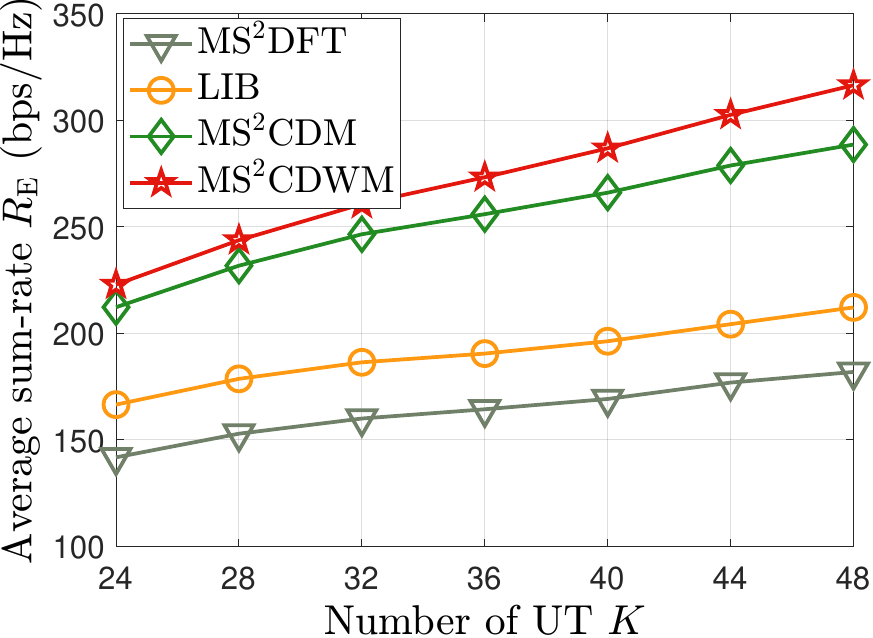}
        \caption{$M_k=3$.}
        \label{fig:sub2}
    \end{subfigure}
    \caption{Average sum rate vs $K$, $S_k=3$, $P_{
    \rm T}=30{\rm dBm}$.}  
 \label{fig M K}
 \vspace{-5mm}
\end{figure}

\begin{figure}[!t]  
    \centering
    \captionsetup{font=footnotesize}
    \begin{subfigure}{0.24\textwidth}  
        \centering
        \includegraphics[width=\linewidth]{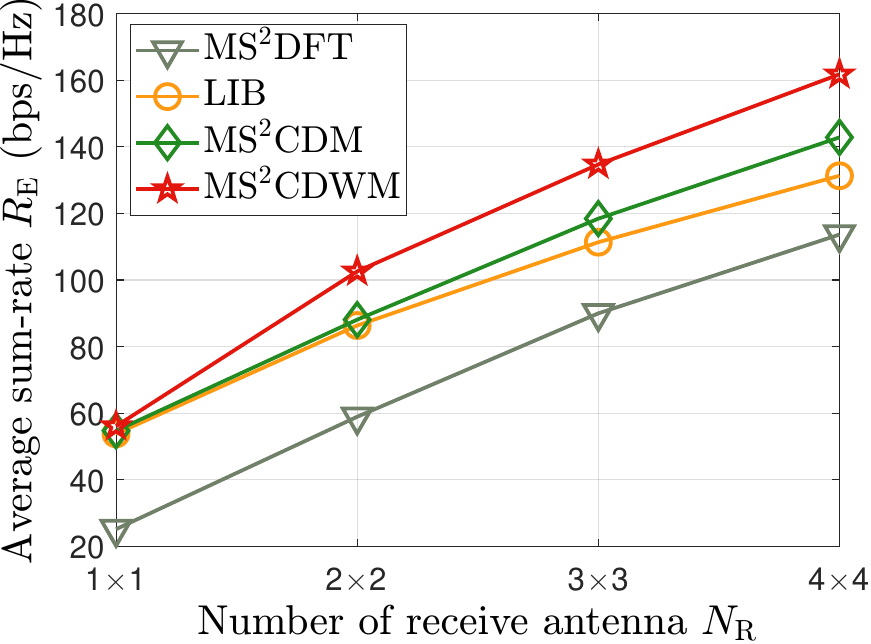}  
        \caption{$P_{
    \rm T}=20{\rm dBm}$}  
        \label{fig:sub1}  
    \end{subfigure}
    \hfill  
    \begin{subfigure}{0.24\textwidth}
        \centering
\includegraphics[width=\linewidth]{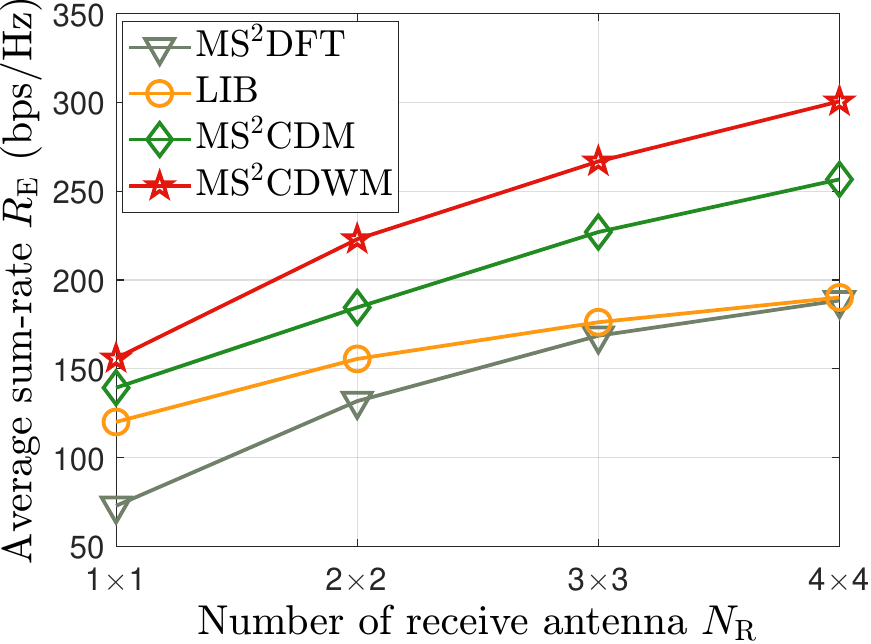}
        \caption{$P_{
    \rm T}=30{\rm dBm}$}
        \label{fig:sub2}
    \end{subfigure}
    \caption{Average sum rate vs $N_{\rm R}$, $S_k=3$, $M_k=1$.}  
    \label{fig NR}  
    \vspace{-5mm}
\end{figure}

\figref{fig M K} compares the impact of UT count on average sum rate. $\text{MS}^2$CDM and $\text{MS}^2$CDWM exhibit faster sum rate growth with increasing UT numbers, attributed to their interference suppression in the beam domain. $\text{MS}^2$CDWM achieves greater sum rate improvement due to further optimization of auxiliary variables $\{{\bf C}_k\}$ and $\{{\bf D}_k\}$. Although the sum rate increases with UT count, per-user average sum rate performance may not, indicating that the number of UTs should be selected based on requirements. \figref{fig NR} evaluates the impact of receive antennas on performance in coherent single-stream transmission. Despite using multi-satellite single-stream transmission, increasing the number of receive antennas significantly enhances performance, driven by improved link budget, which account for different performance gains at low and high signal-to-noise ratios.


\begin{figure}[!t]  
    \centering
    \captionsetup{font=footnotesize}
    \begin{subfigure}{0.24\textwidth}  
        \centering
        \includegraphics[width=\linewidth]{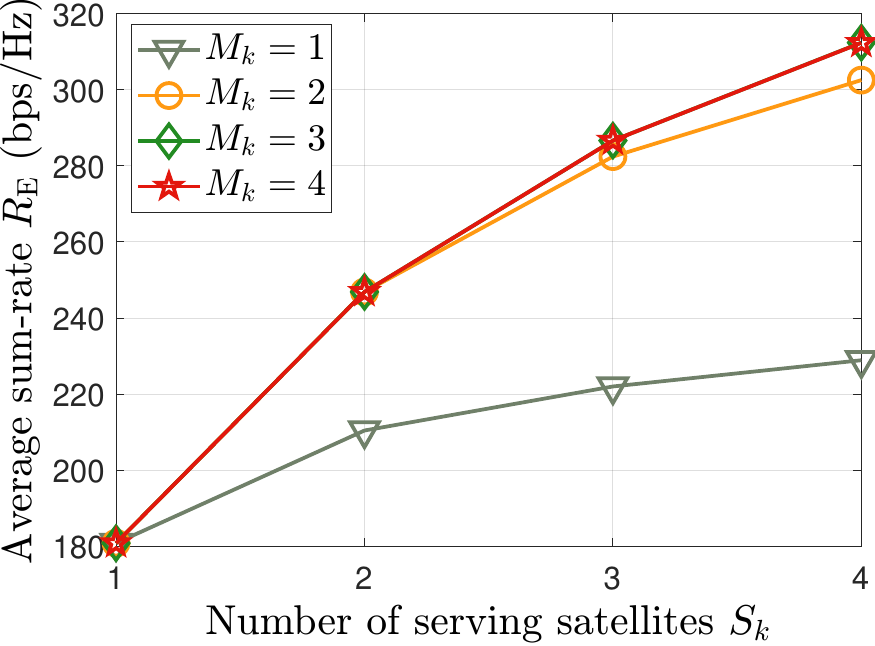}  
        \caption{$B_s =K_s$ (Min. Beams).}  
    \end{subfigure}
    \hfill  
    \begin{subfigure}{0.24\textwidth}
    \centering
\includegraphics[width=\linewidth]{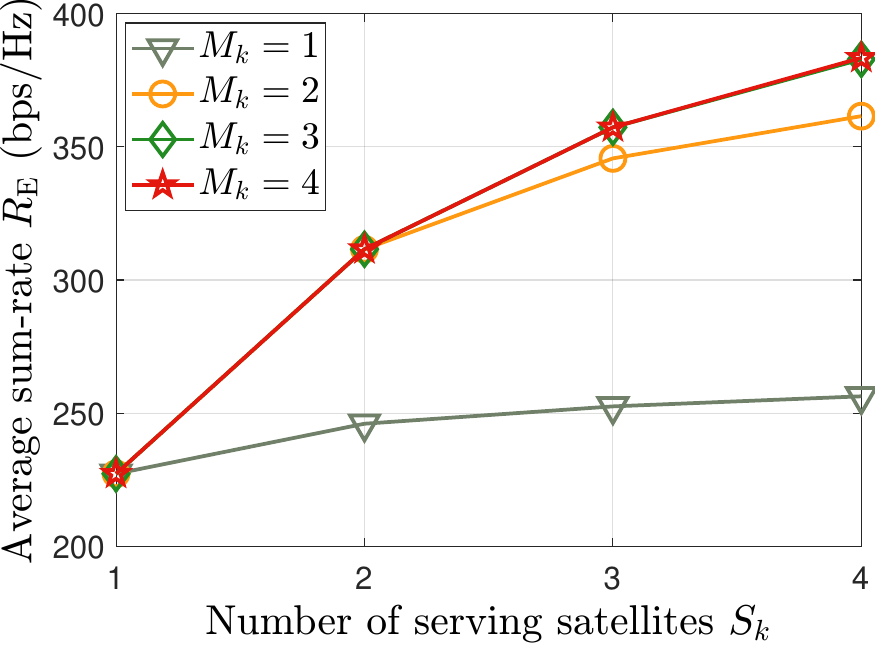}
    \caption{Full channel.}
    \end{subfigure}
    \caption{Average sum rate of $\text{MS}^2$CDWM vs $S_k$ under different $M_k$, $\zeta^2_{s,k} = 0$.}  
    \label{fig M S}  
    \vspace{-6mm}
\end{figure}

\begin{figure}[!t]  
    \centering
    \captionsetup{font=footnotesize}
    \begin{subfigure}{0.24\textwidth}  
    \centering
    \includegraphics[width=\linewidth]{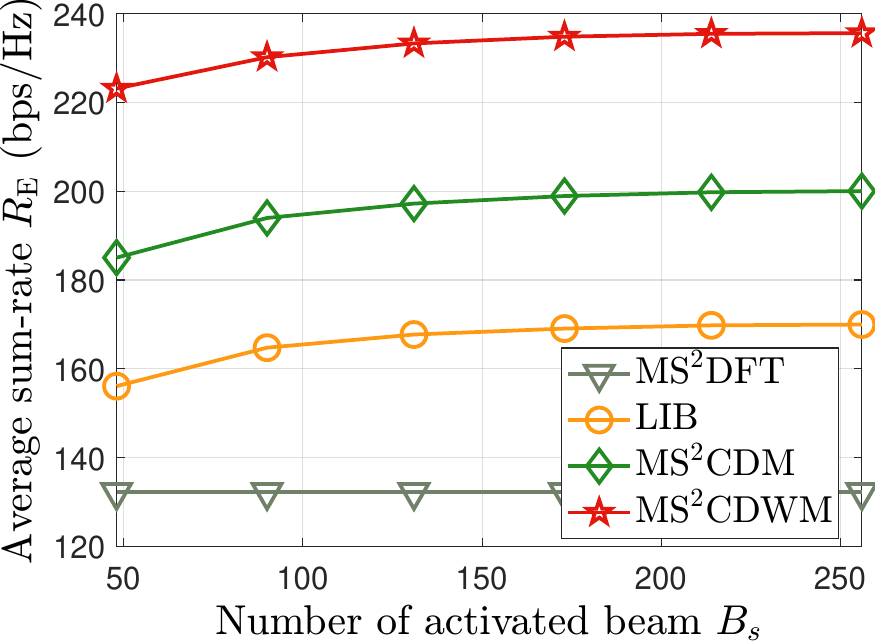}  
        \caption{$M_k=1$.}  
    \end{subfigure}
    \hfill  
    \begin{subfigure}{0.24\textwidth}
    \centering
\includegraphics[width=\linewidth]{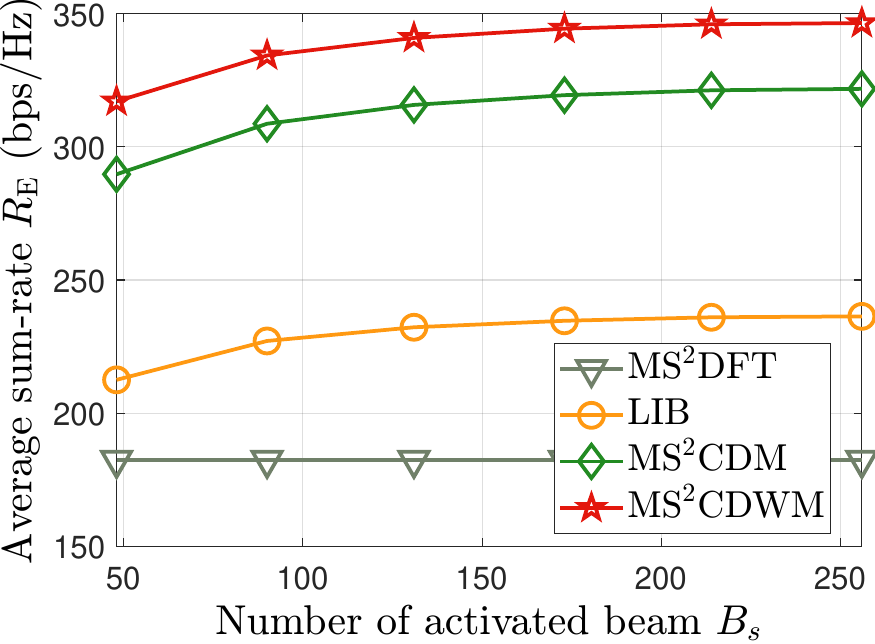}
        \caption{$M_k=3$.}
    \end{subfigure}
    \caption{Average sum rate vs $B_s$, $S_k=3$, $P_{
    \rm T}=30{\rm dBm}$.}  
    \label{fig beam}  
    \vspace{-4mm}
\end{figure}

\figref{fig M S} illustrates the impact of varying numbers of served UTs and data streams under perfect phase compensation. With a fixed number of data streams, increasing the number of serving satellites yields limited performance gains. This is because, under the user-centric satellite clustering, the power resources of cooperating satellites are fully utilized, and adding more satellites in single-stream scenarios only marginally enhances spatial multiplexing capabilities. However, increasing the number of serving satellites raises the rank of each UT’s multi-satellite channel, enabling multi-stream transmission and significantly improving data rates. Due to limited UT reception capabilities, this gain diminishes as the number of data streams increases. \figref{fig beam} compares the performance gains from varying the number of selected beams in beamspace. As the number of beams increases, the channel’s degrees of freedom, power, and inter-user orthogonality improve, driving the performance gain. By analyzing the impact of beam count on performance, the number of beams can be selected based on engineering requirements, balancing computational complexity, cost, implementation challenges, and performance needs.

\begin{figure}[!t]
    \centering
    \captionsetup{font=footnotesize}
    \includegraphics[width=3.5in]{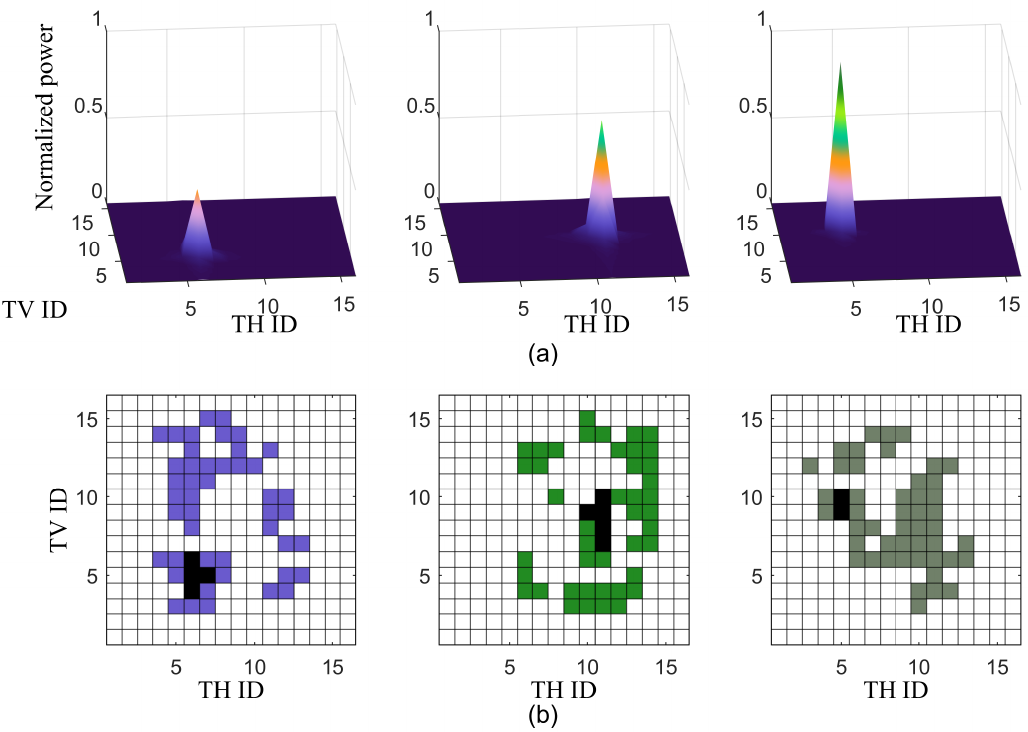}
    \caption{(a) Normalized power of the preselection beam-domain channels between UT $k$ and its serving satellites and (b) beam selections of serving satellites.}
    \label{beam channel fig}
    \vspace{-7mm}
\end{figure}


\figref{beam channel fig}  visualizes multi-satellite beamspace transmission. In (a), the normalized power maps of the preselection beam-domain channels ${\bf H}_{s,k}{\bf F}_s,\ s\in\mathcal{S}_k$ between UT $k$ and its three serving satellites are shown, where the $x$- and $y$-axes denote the horizontal and vertical antenna indices of the UPA, respectively. The beam-domain channels are highly sparse, which supports the applicability of beamspace transmission. Moreover, although each satellite exhibits a single-path characteristic, the multi-satellite channel provides three distinct propagation paths, enabling multi-stream transmission.
In (b), the beams selected at the three serving satellites are highlighted in color, which determines the dimension $B_s$ of beam-domain channel ${\bar {\bf H}}_{s,k}\triangleq{\bf H}_{s,k}{\bf F}_s{\bf A}_s\in\mathbb{C}^{N_{\rm R}\times  B_s}$. The black region, projected from (a), shows the dominant beam-domain power distribution for UT $k$ at each satellite. Only a small subset of beams is selected, which reduces the beam-domain processing dimension while effectively capturing the UT beam-domain channel power.

\vspace{-4mm}
\section{Conclusion}\label{conclusion sec}
\vspace{-1mm}
This paper investigated downlink MSMS beamspace massive MIMO transmission for multi-antenna UTs. This work formulated, for the first time, a signal model for coherent distributed beamspace MIMO that supports multiple data streams transmission, accounting for synchronization errors and revealing new channel characteristics that include multi-rank structures. Based on this model, we established an optimization problem that jointly optimizes satellite clustering, beam selection, and transmit precoding, where the expected sum rate is approximated by its upper bound. For beamspace precoding, we derive an noval equivalent CDWMMSE optimization problem under MSMS transmission, proposed a novel covariance decomposition method, and derived an iterative precoding algorithm based on sCSI. To circumvent the computational burden of iterative processing, we further design several heuristic closed-form precoders.
For satellite clustering, we adopted a competition-based method driven by sCSI and incorporate a redundancy removal module. Furthermore, by exploiting the characteristics of multi-satellite channels, we designed a low-complexity beam selection algorithm focused on enhancing the effective channel power. Simulation results demonstrated that the proposed MSMS beamspace MIMO transmission framework achieves an excellent balance between performance and computational complexity.

\appendices

\vspace{-3mm}
\section{Proof of Proposition \ref{ppn CDWMMSE problem}}\label{pf CDWMMSE problem}
\vspace{-1mm}
By taking the partial derivatives of the objective function of optimization problem $({\text{P3}})$ with respect to $\{{\bf C}_k\}$ and $\{{\bf D}_k\}$ and setting them to zero, we obtain the the optimal expressions: ${\bf D}^{\star}_k = ({\bar {\bf R}}_{{\rm sig}, k}+{\bar {\bf R}}_{{\rm other}, k} )^{-1}{\bar {\bf R}}_{{\rm sig}, k}^{\frac{1}{2}}$ and ${\bf C}^{\star}_k = {\bf E}^{-1}_k,\ \forall k\in\mathcal{K}$.
The validity of these two equations is independent of the value of the other variable.
Substituting these expressions into the objective function of $({\text{P3}})$ yields the following:
\begin{align}
    &\textstyle{\sum_{\forall k}}\beta_k\left[{\rm Tr}({\bf C}_k{\bf E}_k)-\log\det({\bf C}_k)\right]\notag\\
    &\textstyle={\sum_{\forall k}}\beta_k\left[{\rm Tr}({\bf I})-\log\det\left([{\bf I}-({\bar {\bf R}}_{{\rm sig}, k}^{\frac{1}{2}})^H\right.\right.\notag\\
    &\textstyle\qquad\qquad\qquad\left.\left.({\bar {\bf R}}_{{\rm sig}, k}+{\bar {\bf R}}_{{\rm other}, k} )^{-1}{\bar {\bf R}}_{{\rm sig}, k}^{\frac{1}{2}}]^{-1}\right)\right]\notag\\
    &\textstyle={\sum_{\forall k}}\beta_k\left[{\rm Tr}({\bf I})-\log\det({\bf I}+({\bar {\bf R}}_{{\rm sig}, k}^{\frac{1}{2}})^H{\bar {\bf R}}_{{\rm other}, k}^{-1}{\bar {\bf R}}_{{\rm sig}, k}^{\frac{1}{2}})\right]\notag\\
    &\textstyle={\sum_{\forall k}}\beta_k\left[{\rm Tr}({\bf I})-{\bar R}_k\right].
\end{align}
Minimizing this function is equivalent to maximizing the objective function of $({\text{P3}})$. This completes the proof.

\vspace{-2mm}
\section{Proof of Lemma \ref{lemma R half}}\label{pf R half}
\vspace{-1mm}
Based on the constructed expression:
\begin{align}
&\textstyle{\bar {\bf R}}^{\frac{1}{2}}_{{\rm sig}, k}
=[\sum_{s=1}^S{\bar \varphi}_{s,k}\sqrt{\rho_{s,k}}{\bf u}_{s,k}{\bf q}^H_{s,k,k}\quad {\tilde {\boldsymbol{\Sigma}}}_{k}{\bf Q}^H_{k}],
\end{align}
we can derive that
\begin{align}
    &{\bar {\bf R}}^{\frac{1}{2}}_{{\rm sig}, k}({\bar {\bf R}}^{\frac{1}{2}}_{{\rm sig}, k})^H \notag\\
    &= \left(\sum\limits_{s=1}^S{\bar \varphi}_{s,k}\sqrt{\rho_{s,k}}{\bf u}_{s,k}{\bf q}^H_{s,k,k}\right)\left(\sum\limits_{s=1}^S{\bar \varphi}_{s,k}\sqrt{\rho_{s,k}}{\bf u}_{s,k}{\bf q}^H_{s,k,k}\right)^H\notag\\
    &\textstyle\qquad\qquad+\sum_{s=1}^S{\tilde {\boldsymbol{\Sigma}}}_{s,k}({\bf I}_{N_{\rm R}}\otimes{\bf q}^H_{s,k,k})({\bf I}_{N_{\rm R}}\otimes{\bf q}_{s,k,k}){\tilde {\boldsymbol{\Sigma}}}^H_{s,k}\notag\\
    &= \sum\limits_{s_1=1}^S\sum\limits_{s_2=1}^S({\bf q}^H_{s_1,k,k}{\bf q}_{s_2,k,k}){\bar \varphi}_{s_1,k}{\bar \varphi}^H_{s_2,k}\sqrt{\rho_{s_1,k}\rho_{s_2,k}}{\bf u}_{s_1,k}{\bf u}^H_{s_2,k}\notag\\
    &\qquad+{\bf q}^H_{s,k,k}{\bf q}_{s,k,k}[(1\!-\!{\bar \varphi}_{s,k}{\bar \varphi}^H_{s,k})\rho_{s,k}{\bf u}_{s,k}{\bf u}^H_{s,k}+{\tilde \rho}_{s,k}{\boldsymbol{\Sigma}}_{s,k}]
    \notag\\
    &= \sum_{s_1\in\mathcal{S}}\sum_{s_2\in\mathcal{S}}({\bf q}^H_{s_1,k,k}{\bf q}_{s_2,k,k}){\boldsymbol{\Delta}}_{s_1, s_2, k} = {\bar {\bf R}}_{{\rm sig}, k}.
\end{align}
This completes the proof.

\vspace{-3mm}
\section{Proof of Lemma \ref{lemma gradient part1}}\label{pf gradient part1}
\vspace{-1mm}
Denote $r_k={\rm Tr}\left({\bf C}_k{\bf D}^H_k{\bar {\bf R}}_{{\rm sig}, k}{\bf D}_k\right)$, then we have
\begin{align}
    &\textstyle{\rm d}r_k = {\rm Tr}\{{\bf D}_k{\bf C}_k{\bf D}^H_k \textstyle \sum_{\forall (s_1,s_2)}{\rm d}({\bf b}^H_{s_1,k}{\bf b}_{s_2,k}){\boldsymbol{\Delta}}_{s_1, s_2, k}\}  \nonumber \\
    &= \textstyle \sum_{\forall (s_1,s_2)}{\rm Tr}\left\{{\bf D}_k{\bf C}_k{\bf D}^H_k{\boldsymbol{\Delta}}_{s_1, s_2, k}\right\}{\rm d}({\bf b}^H_{s_1,k}{\bf b}_{s_2,k})  \nonumber\\
    &=\textstyle \sum_{\forall (s_1,s_2)}{\psi}_{s_2, s_1,k}{\rm Tr}\left\{{\bf W}^T_{s_1, k}{\bar {\bf v}}_{s_1,k}{\bar {\bf v}}^H_{s_2,k}{\rm d}({\bf W}^*_{s_2,k}) \right\} o_{s_1,k}o_{s_2,k} ,  \nonumber
\end{align}
where ${\psi}_{s_2, s_1,k}={\rm Tr}\left\{{\bf D}_k{\bf C}_k{\bf D}^H_k{\boldsymbol{\Delta}}_{s_1, s_2, k}\right\}$.
Then, we have
\begin{align}
    \textstyle\frac{\partial \sum_{k=1}^Kr_k}{\partial {\bf W}^*_{s,k}}= \sum_{i\in\mathcal{S}}{\bar {\bf h}}^*_{s,k}{\bar {\bf h}}^T_{i,k}{\bf W}_{i, k}  o_{i,k}o_{s,k} {\psi}_{s, i,k}
\end{align}
and $\frac{\partial \sum_{k=1}^K{\rm Tr}\left({\bf C}_k{\bf D}^H_k{\bar {\bf R}}_{{\rm sig}, k}{\bf D}_k\right)}{\partial {\bf W}^*_{k}} = {\boldsymbol{\Xi}}_k{\bf W}_{k}$.

\vspace{-3mm}
\section{Proof of Lemma \ref{lemma gradient part2}}\label{pf gradient part2}
\vspace{-1mm}

\vspace{-3mm}
\begin{align}
&{\rm Tr}\left({\bf C}_k({\bar {\bf R}}_{{\rm sig}, k}^{\frac{1}{2}})^H{\bf D}_k\right) \notag \\
&\textstyle= \sum_{s=1}^S{\rm Tr}\left({\bar \varphi}^H_{s,k}\sqrt{\rho_{s,k}}{\bf D}_k{\bar {\bf C}}_k{\bf q}_{s,k,k}{\bf u}^H_{s,k}\right) \notag\\
&\textstyle\qquad\qquad+ \sum_{s=1}^S{\rm Tr}\left({\bf D}_k{\tilde {\bf C}}_{s,k}{\bf Q}_{s,k}{\tilde {\boldsymbol{\Sigma}}}^H_{s,k}\right)\notag\\
&\textstyle= \sum_{s=1}^S{\rm Tr}\left({\bar \varphi}^H_{s,k}\sqrt{\rho_{s,k}}{\bf D}_k{\bar {\bf C}}_k{\bf W}_{s,k}^H{\bar {\bf v}}^{*}_{s,k}{\bf u}^H_{s,k}o_{s,k}\right) \notag\\
&\textstyle\qquad\qquad+ \sum_{s=1}^S {\rm Tr}({\tilde {\boldsymbol{\Sigma}}}^H_{s,k}{\bf D}_k{\tilde {\bf C}}_{s,k}[{\bf I}_{N_{\rm R}}\otimes{\bf W}^H_{s,k}{\bar{\bf v}}_k^{*})]o_{s,k} \notag \\
&\textstyle=\sum_{s=1}^S{\bar \varphi}^H_{s,k}\sqrt{\rho_{s,k}}{\bf u}^H_{s,k}{\bf D}_k{\bar {\bf C}}_k{\bf W}_{s,k}^H{\bar {\bf v}}^{*}_{s,k}\notag\\
&\textstyle\qquad\qquad+\sum_{s=1}^S\sum_{n=1}^{N_{\rm R}}{\boldsymbol{\sigma}}^H_{s,k,n}{\bf U}^H_k{\bf D}_k{\tilde {\bf C}}_{s,k,n}{\bf W}^H_{s,k}{\bar {\bf v}}^{*}_{s,k}\notag\\
&\textstyle= \sum_{s=1}^S{\bf t}^T_{s,k}{\bf W}_{s,k}^H{\bar {\bf v}}^{*}_{s,k}.
\end{align}
Then we have $\textstyle\frac{\partial {\rm Tr}\left({\bf C}_k({\bar {\bf R}}_{{\rm sig}, k}^{\frac{1}{2}})^H{\bf D}_k\right)}{\partial {\bf W}^*_{k}} = {\breve {\bf V}}^H_k{\bf T}_{k}$.
This completes the proof.

\vspace{-2mm}
\section{Proof of Proposition \ref{ppn optimal solution}}\label{pf optimal solution}
\vspace{-1mm}

The Lagrangian function for  (P5) is given by
\begin{align}
    g_k &\textstyle= {\rm Tr}\left\{\frac{{\boldsymbol{\Upsilon}}_{k}}{\eta_k^2}\!-\! \frac{\beta_k}{\eta_k}{\bf C}_k\!\left({\bf D}^H_k{\bar {\bf R}}_{{\rm sig}, k}^{\frac{1}{2}}\!+\!({\bar {\bf R}}_{{\rm sig}, k}^{\frac{1}{2}})^H{\bf D}_k\right)\!\right\} \notag\\
    &\qquad\qquad+ \lambda_k[{\rm Tr}\left({\bf W}_k{\bf W}^H_k\right)-{\tilde P}_k].
\end{align}
If the optimum are achieved, the derivation with respect to ${\bf W}_k$ and $\eta_k$ must be vanished. According to \lmref{lemma R half}, \lmref{lemma gradient part1}, and \lmref{lemma gradient part2}, we have
\begin{align}
    &\textstyle\frac{\partial g_k}{\partial{\bf W}^{*}_k} = \frac{1}{\eta^2_k}{\boldsymbol{\Xi}}_k{\bf W}_{k}-\frac{\beta_k}{\eta_k}{\breve {\bf V}}^H_k{\bf T}_{k} + \lambda_k{\bf W}_k.
\end{align}
Then we can obtain the following optimal expression
\begin{align}
    {\bf W}^{\star}_k = \eta^{\star}_k(\lambda'_k){\bar {\bf W}}_k,\ {\bar {\bf W}}_k=\left({\boldsymbol{\Xi}}_k+ \lambda'_k{\bf I}\right)^{-1}{\breve {\bf V}}^H_k{\bf T}_{k},
\end{align}
where $\eta^{\star}_k = \sqrt{\frac{{\tilde P}_k}{\|{\bar {\bf W}_k}\|^2_F}}$ and $\lambda'_k = \lambda_k\eta_k^2$. Based on the equation, the Lagrangian function is transformed into a function that only depends on $\lambda'_k$.
Subsequently, by computing the gradient of the Lagrangian function with respect to $\lambda'_k$ via total differentiation and setting it to zero, we obtain the following expression that achieves the optimal value $\textstyle(\lambda'_k)^\star=\frac{\beta_k\sigma_k^2}{P_k}{\rm Tr}\{{\bf D}_k{\bf C}_k{\bf D}^H_k\}$. This completes the proof.

\bibliographystyle{IEEEtran}
\bibliography{reference}

@IEEEtranBSTCTL{IEEEexample:BSTcontrol,
  CTLuse_article_number = "yes",
  CTLuse_paper = "yes",
  CTLuse_forced_etal = "yes",
  CTLmax_names_forced_etal = "4",
  CTLnames_show_etal = "4",
  CTLuse_alt_spacing = "yes",
  CTLalt_stretch_factor = "4",
  CTLdash_repeated_names = "yes",
  CTLname_format_string = "{f.~}{vv~}{ll}{, jj}",
  CTLname_latex_cmd = "",
  CTLname_url_prefix = "[Online]. Available:"
 }

@article{you2020massive,
  author  = {Li You and Ke-Xin Li and Jiaheng Wang and Xiqi Gao and Xiang-Gen Xia and Bj{\"o}rn Ottersten},
  title   = {Massive {MIMO} transmission for {LEO} satellite communications},
  journal = {IEEE J. Sel. Areas Commun.},
  year    = {2020},
  volume  = {38},
  number  = {8},
  pages   = {1851--1865},
  month   = {Aug.}
}

@article{8353925,
  author  = {Miguel {\'A}ngel V{\'a}zquez and M. R. Bhavani Shankar and Charilaos I. Kourogiorgas 
             and Pantelis-Daniel Arapoglou and Vincenzo Icolari and Symeon Chatzinotas 
             and Athanasios D. Panagopoulos and Ana I. P{\'e}rez-Neira},
  title   = {Precoding, Scheduling, and Link Adaptation in Mobile Interactive Multibeam Satellite Systems},
  journal = {IEEE J. Sel. Areas Commun.},
  year    = {2018},
  volume  = {36},
  number  = {5},
  pages   = {971--980},
  month   = {May},
  doi     = {10.1109/JSAC.2018.2832778}
}

@article{8629918,
  author  = {Xin Zhang and Jingjing Wang and Chunxiao Jiang and Chaoxing Yan and Yong Ren and Lajos Hanzo},
  title   = {Robust Beamforming for Multibeam Satellite Communication in the Face of Phase Perturbations},
  journal = {IEEE Trans. Veh. Technol.},
  year    = {2019},
  volume  = {68},
  number  = {3},
  pages   = {3043--3047},
  month   = {Mar.},
  doi     = {10.1109/TVT.2019.2896245}
}

@ARTICLE{9939157,
  author  = {M. Y. Abdelsadek and G. K. Kurt and H. Yanikomeroglu},
  title   = {Distributed Massive {MIMO} for {LEO} Satellite Networks},
  journal = {IEEE Open J. Commun. Soc.},
  volume  = {3},
  pages   = {2162--2177},
  month   = {Nov.},
  year    = {2022},
  doi     = {10.1109/OJCOMS.2022.3219419}
}

@Article{10380500,
  author    = {X. Zhang and S. Sun and M. Tao and Q. Huang and X. Tang},
  title     = {Multi-Satellite Cooperative Networks: Joint Hybrid Beamforming and User Scheduling Design},
  journal   = {IEEE Trans. Wireless Commun.},
  year      = {2024},
  volume    = {23},
  number    = {7},
  pages     = {7938--7952},
  month     = {Jul.},
  doi       = {10.1109/TWC.2023.3346463},
  publisher = {IEEE},
  issn      = {1536-1276}
}

@article{shi2011iteratively,
  author  = {Shi, Qingjiang and Razaviyayn, Meisam and Luo, Zhi-Quan and He, Chen},
  title   = {An iteratively weighted {MMSE} approach to distributed sum-utility maximization for a {MIMO} interfering broadcast channel},
  journal = {IEEE Trans. Signal Process.},
  volume  = {59},
  number  = {9},
  pages   = {4331--4340},
  year    = {2011},
  month = {Sept.}
}

@article{shen2018fractional,
  author  = {Kaiming Shen and Wei Yu},
  title   = {Fractional programming for communication systems—Part {I}: Power control and beamforming},
  journal = {IEEE Trans. Signal Process.},
  year    = {2018},
  volume  = {66},
  number  = {10},
  pages   = {2616--2630},
  month   = {Oct.},
  doi     = {10.1109/TSP.2018.2828818}
}

@article{wang2021resource,
  author  = {Wenjin Wang and Linna Gao and Rui Ding and Jizhao Lei and Li You and Chien Aun Chan and Xiqi Gao},
  title   = {Resource efficiency optimization for robust beamforming in multi-beam satellite communications},
  journal = {IEEE Trans. Veh. Technol.},
  year    = {2021},
  volume  = {70},
  number  = {7},
  pages   = {6958--6968},
  month   = {Jul.}
}

@inproceedings{10008605,
  author    = {V. N. Ha and Z. Abdullah and G. Eappen and J. C. M. Duncan and R. Palisetty and 
               J. L. G. Rios and W. A. Martins and H.-F. Chou and J. A. Vasquez and 
               L. M. Garces-Socarras and H. Chaker and S. Chatzinotas},
  title     = {Joint Linear Precoding and {DFT} Beamforming Design for Massive {MIMO} Satellite Communication},
  booktitle = {Proc. IEEE Global Commun. Conf. (GLOBECOM)},
  address   = {Rio de Janeiro, Brazil},
  month     = {Dec.},
  year      = {2022},
  pages     = {1121--1126},
  doi       = {10.1109/GCWkshps56602.2022.10008605}
}

@ARTICLE{li2021downlink,
  author  = {K.-X. Li and L. You and J. Wang and X. Gao and C. G. Tsinos and S. Chatzinotas and B. Ottersten},
  title   = {Downlink Transmit Design for Massive {MIMO} {LEO} Satellite Communications},
  journal = {IEEE Trans. Commun.},
  volume  = {70},
  number  = {2},
  pages   = {1014--1028},
  month   = {Feb.},
  year    = {2021},
  doi={10.1109/TCOMM.2021.3131573}
}

@article{liu2025time,
  author  = {Liu, Zijun and Wang, Yafei and Fang, Tianhao and Wang, Wenjin and Sun, Zhili},
  title   = {Time-Continuous Frequency Allocation for Feeder Links of Mega Constellations with Multi-Antenna Gateway Stations},
  journal = {arXiv},
  year    = {2025},
  month   = {May.},
  doi     = {10.48550/arXiv.2505.12429},
  url     = {https://arxiv.org/abs/2505.12429}
}

@article{Wang2025DP,
  author  = {Wang, Yafei and Ha, Vu Nguyen and Ntontin, Konstantinos and Yan, Hong and Wang, Wenjin and Chatzinotas, Symeon and Ottersten, Bj{\"o}rn},
  title   = {Statistical {{CSI}}-Based Distributed Precoding Design for {{OFDM}}-Cooperative Multi-Satellite Systems},
  journal = {arXiv},
  year    = {2025},
  month   = {May.},
  doi     = {10.48550/arXiv.2505.08038},
  url     = {https://arxiv.org/abs/2505.08038}
}

@inproceedings{zhu2025downlink,
  author    = {Feng Zhu and Yunfei Wang and Xiqi Gao},
  title     = {Downlink Precoding for Multi-Beam {LEO} Satellite Communications with Asynchronous Interference},
  booktitle = {Proc. IEEE Wireless Commun. Networking Conf. (WCNC)},
  address   = {Milan, Italy},
  month     = {Mar.},
  year      = {2025},
  pages     = {1--6}
}

@article{christensen2008weighted,
  author  = {Christensen, Søren Skovgaard and Agarwal, Rajiv and De Carvalho, Elisabeth and Cioffi, John M.},
  title   = {Weighted sum-rate maximization using weighted {MMSE} for {MIMO-BC} beamforming design},
  journal = {IEEE Trans. Wireless Commun.},
  volume  = {7},
  number  = {12},
  pages   = {4792--4799},
  year    = {2008},
month = {Dec.}
}

@article{9174860,
  author  = {Shuaifei Chen and Jiayi Zhang and Emil Bj{\"o}rnson and Jing Zhang and Bo Ai},
  title   = {Structured Massive Access for Scalable Cell-Free Massive {MIMO} Systems},
  journal = {IEEE J. Sel. Areas Commun.},
  year    = {2021},
  volume  = {39},
  number  = {4},
  pages   = {1086--1100},
  month   = {Apr.},
  doi     = {10.1109/JSAC.2020.3018836}
}

@ARTICLE{hou2024joint,
  author  = {H. Hou and Y. Wang and X. Yi and W. Wang and S. Jin},
  title   = {Joint Beam Alignment and Doppler Estimation for Fast Time-Varying Wideband {mmWave} Channels},
  journal = {IEEE Trans. Wireless Commun.},
  volume  = {23},
  number  = {9},
  pages   = {10895--10910},
  month   = {Sept.},
  year    = {2024}
}

@article{de2025applicability,
  author  = {Riccardo De Gaudenzi and Giacomo Bacci and Marco Luise and Luca Sanguinetti and Piero Angeletti},
  title   = {Applicability of {CF}-{MIMO} Precoding to a Formation of Arrays ({FoA}) for Mobile Satellite Communications},
  journal = {IEEE Trans. Aerosp. Electron. Syst.},
  volume  = {61},
  number  = {5},
  pages   = {11069--11087},
  month   = {Oct.},
  year    = {2025},
  doi     = {10.1109/TAES.2025.3556662}
}

@article{liu2025gcn,
  author    = {Liu, Zhilong and Zhang, Jiayi and Xu, Bokai and Ng, Derrick Wing Kwan and Nallanathan, Arumugam and Ai, Bo},
  title     = {{{GCN}}-based Low-Complexity Downlink Beamforming for Cell-Free Massive {{MIMO}} Systems with Partially Coherent Joint Transmission},
  journal   = {IEEE Trans. Wireless Commun.},
  year      = {2025},
  volume    = {24},
  number    = {12},
  pages     = {10440--10455},
  month     = {Dec.},
  doi       = {10.1109/TWC.2025.3579666},
  publisher = {IEEE},
  issn      = {1536-1276}
}

@techreport{3GPP_TR_36_819,
  author       = {3GPP},
  title        = {{TR} 36.819 V11.2.0: Coordinated multi-point operation for {LTE} physical layer aspects (Release 11)},
  institution  = {3GPP},
  type         = {Tech. Rep.},
  number       = {TR 36.819 V11.2.0},
  year         = {2013},
  month        = {Sep.}
}

@inproceedings{6902008,
  author    = {Burkhardt, F. and Jaeckel, S. and Eberlein, E. and Prieto-Cerdeira, R.},
  title     = {{QuaDRiGa}: A {MIMO} Channel Model for Land Mobile Satellite},
  booktitle = {Proc. 8th European Conf. Antennas Propag. (EuCAP 2014)},
  year      = {2014},
    month = {Apr.},
  pages     = {1274--1278},
  doi       = {10.1109/EuCAP.2014.6902008}
}

@inproceedings{9815679,
  author    = {Stephan Jaeckel and Leszek Raschkowski and Lars Thieley},
  title     = {A {5G-NR} Satellite Extension for the {QuaDRiGa} Channel Model},
  booktitle = {Proc. Joint Eur. Conf. Netw. Commun. \& 6G Summit (EuCNC/6G Summit)},
  pages     = {142--147},
  year      = {2022},
  doi       = {10.1109/EuCNC/6GSummit54941.2022.9815679}
}

@manual{QuaDRiGa2023,
  author       = {{Fraunhofer Heinrich Hertz Institute}},
  title        = {Quasi Deterministic Radio Channel Generator: User Manual and Documentation},
  institution  = {Fraunhofer Heinrich Hertz Institute, Wireless Communications and Networks},
  edition      = {v2.8.1},
  address      = {Einsteinufer 37, 10587 Berlin, Germany},
  year         = {2023},
  month        = {Dec.},
  note         = {Available: \url{https://github.com/fraunhoferhhi/QuaDRiGa}},
  url          = {http://www.quadriga-channel-model.de}
}

@techreport{3GPP_TR_38_811,
  author       = {3GPP},
  title        = {{TR} 38.811 V15.4.0: Study on New Radio ({NR}) to Support Non-Terrestrial Networks},
  institution  = {3GPP},
  type         = {Tech. Rep.},
  number       = {TR 38.811 V15.4.0},
  year         = {2020},
  month        = {Sep.}
}

@article{10596023,
  author  = {Chen, Xin and Luo, Zhiyong},
  title   = {Asynchronous Interference Mitigation for {LEO} Multi-Satellite Cooperative Systems},
  journal = {IEEE Trans. Wireless Commun.},
  month   = {Oct.},
  year    = {2024},
  volume  = {23},
  number  = {10},
  pages   = {14956--14971},
  doi     = {10.1109/TWC.2024.3422101}
}

@techreport{3gpp_tr_38_821,
  author       = {3GPP},
  title        = {{TR} 38.821 V16.2.0: Solutions for {NR} to Support Non-Terrestrial Networks ({NTN})},
  institution  = {3GPP},
  type         = {Tech. Rep.},
  number       = {TR 38.821 V16.2.0},
  year         = {2023},
  month        = {Mar.}
}

@techreport{SpaceX_Gen2_2021,
  title        = {Federal Communications Commission; Amendment to Pending Application for the {SpaceX} {Gen2} {NGSO} Satellite System},
  institution  = {FCC},
  number       = {File No. SAT-AMD-2021},
  type         = {Tech. Rep.},
  month        = {August},
  year         = {2021},
  address      = {Washington, D.C.},
  note         = {Available: https://fcc.report/IBFS/SAT-AMD-20210818-00105/12943361.pdf}
}

@article{9998075,
  author  = {Yu, Li and Wan, Jixiang and Zhang, Kai and Teng, Fei and Lei, Liujie and Liu, Yukai},
  title   = {Spaceborne Multibeam Phased Array Antennas for Satellite Communications},
  journal = {IEEE Aerosp. Electron. Syst. Mag.},
  volume  = {38},
  number  = {3},
  pages   = {28--47},
  year    = {2023},
month = {Mar.},
  doi     = {10.1109/MAES.2022.3231580}
}

@article{10440321,
  author  = {Xiang, Ziyu and Gao, Xiqi and Li, Ke-Xin and Xia, Xiang-Gen},
  title   = {Massive {MIMO} Downlink Transmission for Multiple {LEO} Satellite Communication},
  journal = {IEEE Trans. Commun.},
  month   = {Jun.},
  year    = {2024},
  volume  = {72},
  number  = {6},
  pages   = {3352--3364},
  doi     = {10.1109/TCOMM.2024.3367726}
}

@ARTICLE{Gao2016Near,
  author  = {X. Gao and L. Dai and Z. Chen and Z. Wang and Z. Zhang},
  title   = {Near-optimal beam selection for beamspace {mmWave} massive {MIMO} systems},
  journal = {IEEE Commun. Lett.},
  volume  = {20},
  number  = {5},
  pages   = {1054--1057},
  month   = {May},
  year    = {2016}
}

@article{10820534,
  author  = {Konstantinos Ntontin and Eva Lagunas and Jorge Querol and Junaid ur Rehman and Joel Grotz and Symeon Chatzinotas and Bj{\"o}rn Ottersten},
  title   = {A Vision, Survey, and Roadmap Toward Space Communications in the {6G} and Beyond Era},
  journal = {Proc. IEEE},
  volume  = {},
  number  = {},
  pages   = {1--37},
  month   = {Jan.},
  year    = {2025},
  doi     = {10.1109/JPROC.2024.3512934}
}

@article{WANG2025,
title = {Toward Mobile Satellite Internet: The Fundamental Limitation of Wireless Transmission and Enabling Technologies},
journal = {Engineering},
year = {2025},
issn = {2095-8099},
doi = {https://doi.org/10.1016/j.eng.2025.07.007},
url = {https://www.sciencedirect.com/science/article/pii/S2095809925003698},
author = {Wenjin Wang and Yiming Zhu and Yafei Wang and Rui Ding and Symeon Chatzinotas}
}

@article{bjornson2014optimal,
  title={Optimal Multiuser Transmit Beamforming: A Difficult Problem with a Simple Solution Structure [Lecture Notes]},
  author={Bj{\"o}rnson, Emil and Bengtsson, Mats and Ottersten, Bj{\"o}rn},
  journal={{IEEE Signal Process. Mag.}},
  volume={31},
  number={4},
  pages={142--148},
  month={Jul.},
  year={2014},
  publisher={IEEE}
}

@article{11049893,
  author    = {Wang, Yafei and Hou, Hongwei and Yi, Xinping and Wang, Wenjin and Jin, Shi},
  title     = {Toward Unified {AI} Models for {MU-MIMO} Communications: A Tensor Equivariance Framework},
  journal   = {IEEE Trans. Wireless Commun.},
  year      = {2025},
  volume    = {24},
  number    = {12},
  pages     = {10517--10533},
  month     = {Dec.},
  doi       = {10.1109/TWC.2025.3580321},
  publisher = {IEEE},
  issn      = {1536-1276}
}

@Article{wang2022weighted,
  author  = {Wang, Yafei and Wang, Wenjin and You, Li and Tsinos, Christos G. and Jin, Shi},
  journal = {IEEE Wireless Commun. Lett.},
  title   = {Weighted {MMSE} Precoding for Constructive Interference Region},
  year    = {2022},
  number  = {12},
  pages   = {2605-2609},
  volume  = {11},
  doi     = {10.1109/LWC.2022.3211731},
}

@ARTICLE{Dong2025statistical,
  author  = {Q. Dong and Y. Wang and N. Hu and Y. Zhu and W. Wang and L. Chai},
  title   = {Statistical {CSI}-Based Beamspace Transmission for Massive {MIMO} {LEO} Satellite Communications},
  volume  = {27},
  number  = {12},
  pages   = {1214},
  month   = {Dec.},
  year    = {2025}
}

@inproceedings{10437228,
  author    = {Shiyu Wu and Yafei Wang and Gangle Sun and Li You and Wenjin Wang and Rui Ding},
  title     = {Energy and Computational Efficient Precoding for {LEO} Satellite Communications},
  booktitle = {Proc. IEEE Glob. Commun. Conf. (GLOBECOM)},
  address   = {Kuala Lumpur, Malaysia},
  month     = {Dec.},
  year      = {2023},
  pages     = {1872--1877},
  doi       = {10.1109/GLOBECOM54140.2023.10437228}
}

@ARTICLE{Wu2023Simultaneous,
  author  = {K. Wu and J. A. Zhang and X. Huang and Y. J. Guo and L. Hanzo},
  title   = {Simultaneous beam and user selection for the beamspace {mmWave}/{THz} massive {MIMO} downlink},
  journal = {IEEE Trans. Commun.},
  volume  = {71},
  number  = {3},
  pages   = {1785--1797},
  month   = {Mar.},
  year    = {2023}
}

@article{wu2025distributed,
  title        = {Distributed Beamforming for Multiple {LEO} Satellites With Imperfect Delay and {Doppler} Compensations: Modeling and Rate Analysis},
author={Wu, Shiyu and Wang, Yafei and Sun, Gangle and Wang, Wenjin and Wang, Jiaheng and Ottersten, Björn},
journal      = {IEEE Trans. Veh. Technol.},
    year      = {2025},
  volume    = {74},
  number    = {9},
  pages     = {14978--14984},
  month     = {Sept.},
  doi       = {10.1109/TVT.2025.3564047},
  publisher = {IEEE},
}

@INPROCEEDINGS{ha2024user,
  author    = {V. N. Ha and D. H. N. Nguyen and J. C.-M. Duncan and J. L. Gonzalez-Rios 
               and J. A. V. Peralvo and G. Eappen and L. M. Garces-Socarras 
               and R. Palisetty and S. Chatzinotas and B. Ottersten},
  title     = {User-Centric Beam Selection and Precoding Design for Coordinated Multiple-Satellite Systems},
  booktitle = {Proc. IEEE Int. Symp. Pers., Indoor Mobile Radio Commun. (PIMRC)},
  address   = {Valencia, Spain},
  month     = {Sept.},
  year      = {2024}
}

@INPROCEEDINGS{wang2025DP_Conf,
  author    = {Y. Wang and V. N. Ha and K. Ntontin and W. Wang and S. Chatzinotas and B. Ottersten},
  title     = {Statistical {CSI}-Based Distributed Precoding for Multi-Satellite Cooperative Transmission},
  booktitle = {Proc. IEEE Veh. Technol. Conf. (VTC2025-Fall)},
  address   = {Chengdu, China},
  month     = {Oct.},
  year      = {2025}
}

@inproceedings{Tamiru2025Distributed,
  author    = {B. Tamiru and K. Ntontin and V. N. Ha and S. Chatzinotas},
  title     = {Distributed Precoding Design for Satellite Swarms under Imperfect Phase Synchronization},
  booktitle = {Proc. IEEE Global Commun. Conf. (GLOBECOM)},
  address   = {San Francisco, CA, USA},
  month     = {Dec.},
  year      = {2025}
}

@article{cao2025interference,
  author        = {Cao, Wenjing and Wang, Yafei and Ji, Tianxiang and Cao, Tianyang and Wang, Wenjin and Chatzinotas, Symeon and Ottersten, Bj{\"o}rn},
  title         = {Interference in Spectrum-Sharing Integrated Terrestrial and Satellite Networks: Modeling, Approximation, and Robust Transmit Beamforming},
  journal       = {arXiv},
  year          = {2025},
  month         = {Jun.},
  doi           = {10.48550/arXiv.2506.11851},
  url           = {https://arxiv.org/abs/2506.11851}
}

@inproceedings{9625524,
  author    = {Kisseleff, Steven and Lagunas, Eva and Krivochiza, Jevgenij and Querol, Jorge and Maturo, Nicola and Marrero, Liz Martinez and Merlano-Duncan, Juan and Chatzinotas, Symeon},
  title     = {Centralized Gateway Concept for Precoded Multi-beam {{GEO}} Satellite Networks},
  booktitle = {Proc. IEEE 94th Veh. Technol. Conf. (VTC2021-Fall)},
  year      = {2021},
  month     = {Sep.},
  pages     = {1--6},
  doi       = {10.1109/VTC2021-Fall52928.2021.9625524},
  publisher = {IEEE},
  note      = {Virtual Conf., Sep. 27--28, 2021}
}

@book{tse2005fundamentals,
  title={Fundamentals of wireless communication},
  author={Tse, David and Viswanath, Pramod},
  year={2005},
  publisher={Cambridge university press}
}

@article{zhu2024joint,
  author    = {Zhu, Yiming and Zhuang, Jiawei and Sun, Gangle and Hou, Hongwei and You, Li and Wang, Wenjin},
  title     = {Joint Channel Estimation and Prediction for Massive {MIMO} With Frequency Hopping Sounding},
  journal   = {IEEE Trans. Commun.},
  year      = {2025},
  volume    = {73},
  number    = {7},
  pages     = {5139--5154},
  month     = {Jul.},
  doi       = {10.1109/TCOMM.2024.3523972},
  publisher = {IEEE},
  issn      = {0090-6778}
}

@article{zhang2025decentralized,
  author  = {Zhang, Yuchen and Lagunas, Eva and Zheng, Xue Xian and Chatzinotas, Symeon and Al-Naffouri, Tareq Y.},
  title   = {Decentralized Cooperative Beamforming for Networked {{LEO}} Satellites with Statistical {{CSI}}},
  journal = {arXiv},
  year    = {2025},
  month   = {Dec.},
  doi     = {10.48550/arXiv.2512.18890},
  url     = {https://arxiv.org/abs/2512.18890}
}
	
\end{document}